\documentclass[twocolumn]{aastex63}

\usepackage{amsmath}

\received{\today}
\revised{\today}

\submitjournal{AJ}

\shorttitle{A clock stabilization system for CHIME/FRB Outriggers}
\shortauthors{Mena-Parra et al.}
\graphicspath{{./}{figures/}}

\begin{document}

\title{A clock stabilization system for CHIME/FRB Outriggers}

\correspondingauthor{J.~Mena-Parra}
\email{jdmena@mit.edu}

\author[0000-0002-0772-9326]{J.~Mena-Parra}
\affiliation{MIT Kavli Institute for Astrophysics and Space Research, Massachusetts Institute of Technology, 77 Massachusetts Ave, Cambridge, MA 02139, USA}

\author[0000-0002-4209-7408]{C.~Leung}
\affiliation{MIT Kavli Institute for Astrophysics and Space Research, Massachusetts Institute of Technology, 77 Massachusetts Ave, Cambridge, MA 02139, USA}
\affiliation{Department of Physics, Massachusetts Institute of Technology, 77 Massachusetts Ave, Cambridge, MA 02139, USA}

\author[0000-0003-1860-1632]{S.~Cary}
\affiliation{Department of Astronomy, Wellesley College, 106 Central Street, Wellesley, MA 02481, USA}
\affiliation{MIT Kavli Institute for Astrophysics and Space Research, Massachusetts Institute of Technology, 77 Massachusetts Ave, Cambridge, MA 02139, USA}

\author[0000-0002-4279-6946]{K.~W.~Masui}
\affiliation{MIT Kavli Institute for Astrophysics and Space Research, Massachusetts Institute of Technology, 77 Massachusetts Ave, Cambridge, MA 02139, USA}
\affiliation{Department of Physics, Massachusetts Institute of Technology, 77 Massachusetts Ave, Cambridge, MA 02139, USA}

\author[0000-0003-4810-7803]{J.~F.~Kaczmarek}
\affiliation{National Research Council Canada, Herzberg Astronomy and Astrophysics Research Centre, Dominion Radio Astrophysical Observatory, PO Box 248, Penticton, British Columbia, V2A 6J9 Canada}

\author[0000-0001-6523-9029]{M.~Amiri}
  \affiliation{Department of Physics and Astronomy, University of British Columbia, 6224 Agricultural Road, Vancouver, BC V6T 1Z1 Canada}
 
\author[0000-0003-3772-2798]{K.~Bandura}
  \affiliation{Lane Department of Computer Science and Electrical Engineering, 1220 Evansdale Drive, PO Box 6109, Morgantown, WV 26506, USA}
  \affiliation{Center for Gravitational Waves and Cosmology, West Virginia University, Chestnut Ridge Research Building, Morgantown, WV 26505, USA}
  
\author[0000-0001-8537-9299]{P.~J.~Boyle}
  \affiliation{Department of Physics, McGill University, 3600 rue University, Montr\'eal, QC H3A 2T8, Canada}
  \affiliation{McGill Space Institute, McGill University, 3550 rue University, Montr\'eal, QC H3A 2A7, Canada}
  
\author[0000-0003-2047-5276]{T.~Cassanelli}
  \affiliation{Dunlap Institute for Astronomy \& Astrophysics, University of Toronto, 50 St.~George Street, Toronto, ON M5S 3H4, Canada}
  \affiliation{David A.~Dunlap Department of Astronomy \& Astrophysics, University of Toronto, 50 St.~George Street, Toronto, ON M5S 3H4, Canada}

\author[0000-0001-6509-8430]{J.-F.~Cliche}
  \affiliation{Department of Physics, McGill University, 3600 rue University, Montr\'eal, QC H3A 2T8, Canada}
  
\author[0000-0001-7166-6422]{M.~Dobbs}
  \affiliation{Department of Physics, McGill University, 3600 rue University, Montr\'eal, QC H3A 2T8, Canada}
  \affiliation{McGill Space Institute, McGill University, 3550 rue University, Montr\'eal, QC H3A 2A7, Canada}
 
\author[0000-0001-9345-0307]{V.~M.~Kaspi}
  \affiliation{Department of Physics, McGill University, 3600 rue University, Montr\'eal, QC H3A 2T8, Canada}
  \affiliation{McGill Space Institute, McGill University, 3550 rue University, Montr\'eal, QC H3A 2A7, Canada}
  
\author[0000-0003-1455-2546]{T.~L.~Landecker}
  \affiliation{Dominion Radio Astrophysical Observatory, Herzberg Research Centre for Astronomy and Astrophysics, National Research Council Canada, PO Box 248, Penticton, BC V2A 6J9, Canada}
  
\author[0000-0003-2116-3573]{A.~Lanman}
  \affiliation{Department of Physics, McGill University, 3600 rue University, Montr\'eal, QC H3A 2T8, Canada}
  \affiliation{McGill Space Institute, McGill University, 3550 rue University, Montr\'eal, QC H3A 2A7, Canada}
 
\author[0000-0001-6903-5074]{J.~L.~Sievers}
  \affiliation{McGill Space Institute, McGill University, 3550 rue University, Montr\'eal, QC H3A 2A7, Canada}
  \affiliation{Perimeter Institute for Theoretical Physics, 31 Caroline Street N, Waterloo, ON N25 2YL, Canada}
  \affiliation{School of Chemistry and Physics, University of KwaZulu-Natal, 238 Mazisi Kunene Rd, Glenwood, Durban, 4041, South Africa}
  
\collaboration{99}{(CHIME/FRB Collaboration)}



\begin{abstract}
The Canadian Hydrogen Intensity Mapping Experiment (CHIME) has emerged as
the prime telescope for detecting fast radio bursts (FRBs). 
CHIME/FRB Outriggers will be a dedicated very-long-baseline 
interferometry (VLBI) instrument consisting of outrigger
telescopes at continental baselines working with CHIME 
and its specialized real-time transient-search
backend (CHIME/FRB) to detect and localize FRBs with 50~mas
precision. In this paper we present a minimally invasive clock
stabilization system that effectively transfers the CHIME digital backend
reference clock from its original GPS-disciplined ovenized
crystal oscillator to a passive hydrogen maser. This 
enables us to 
combine the long-term stability and absolute time tagging 
of the GPS clock with the 
short and intermediate-term 
stability of the maser to reduce the 
clock timing errors between VLBI calibration 
observations. We validate the system with VLBI-style observations of 
Cygnus A over a 400~m baseline between CHIME and the CHIME Pathfinder, 
demonstrating agreement between sky-based and maser-based timing measurements
at the $30$~ps rms level on timescales 
ranging from one minute to up to nine days, 
and meeting the stability requirements
for CHIME/FRB Outriggers. In addition, we present an alternate 
reference clock solution for outrigger stations which lack the 
infrastructure to support a passive hydrogen maser.

\end{abstract}

\keywords{Radio astronomy(1338), Radio transient sources (2008), 
Radio pulsars (1353), Astronomical instrumentation(799), 
Very long baseline interferometry (1769)}


\section{Introduction}
\label{sec:intro}

Fast radio bursts \citep[FRBs,][]{lorimer2007bright} are transient pulses 
of radio light observed out to cosmological distances; both their origins
and emission mechanisms remain unclear. Even though thousands of FRB events 
occur over the full sky every day \citep{CHIMEFRB_CAT1,2018MNRAS.475.1427B},
their detection with traditional radio telescopes is challenging due to the 
randomly-occurring nature of the majority of bursts. 

With its unique design optimized for rapid wide-field 
observations and a powerful 
real-time transient-search engine \citep[CHIME/FRB,][]{FRBSystemOverview},
the Canadian Hydrogen Intensity Mapping Experiment 
\citep[CHIME\footnote{\href{https://chime-experiment.ca}
{https://chime-experiment.ca}}, ][]{chime_overview_2021}
has become the leading facility for detection of FRBs,
detecting over 500 FRBs \citep{CHIMEFRB_CAT1} and
18 new repeating sources \citep{R2,RN, RN2}
in its first year of full operations. 
Such an unprecedented sample of events with a single survey has enabled
detailed studies of statistical properties of the FRB population 
such as fluence distribution and sky rate, scattering time,
dispersion measure (DM) distribution, spatial 
distribution, burst morphology, and correlations with large-scale structure
\citep{CHIMEFRB_CAT1,FRB_Morphology2021,FRB_galactic_lat2021,FRB_lss_xcor2021,
2021arXiv210710858C}.

However, except for FRBs with low dispersion measure 
\citep{michilli2020analysis,Bhardwaj_M81,2021arXiv210812122B}, 
CHIME/FRB's arcminute localization 
precision is insufficient to localize these bursts to their host galaxies, 
which is crucial to understand their nature and unlock their 
potential as probes of the intergalactic medium and large-scale structure. 
To overcome this limitation, the CHIME/FRB collaboration is currently 
developing CHIME/FRB Outriggers, a program to deploy 
CHIME-like outrigger telescopes at continental baseline distances.
CHIME and the outriggers will
form a dedicated very-long-baseline interferometry (VLBI) network 
capable of detecting hundreds of FRBs each year with sub-arcsecond
localization precision in near real-time, 
allowing for the unique identification of FRB
galaxy hosts and source environments.

Because VLBI localizes sources by precisely measuring the
difference in the arrival time
 of astronomical signals between independent telescopes across 
far-separated sites, it is critical to use very stable 
local reference signals (i.e., clocks) that allow
the synchronization of VLBI stations without losing coherence during
observations and between calibrations. 
This is particularly important for stationary telescopes like CHIME
and the outrigger stations that can only be calibrated when a bright
radio source transits through their field of view.
The superior stability performance of
hydrogen masers on short and intermediate timescales makes them the 
preferred option for VLBI applications 
\citep{2018PASP..130a5002M,2019ApJ...875L...2E,2020arXiv200409987S}. 
Here, we present a
hardware and software clock stabilization solution for the CHIME 
telescope that effectively
transfers the reference clock from its original GPS-disciplined
crystal oscillator to a passive hydrogen maser during 
VLBI observations, meeting the timing requirements for FRB VLBI with CHIME/FRB
Outriggers. Furthermore, this system can be implemented
without interrupting CHIME's current observational campaign and without
modifications to the correlator or 
the data-analysis pipelines 
for cosmology and radio transient science.

The paper is organized as follows: 
Section~\ref{sec:inst_overview} describes 
the features of the CHIME instrument that are relevant to its use as
a VLBI station in CHIME/FRB Outriggers.
Section~\ref{sec:requirements} discusses the CHIME/FRB Outriggers
clock stability requirements for FRB VLBI. 
Section~\ref{sec:clock_stabilization} describes the hardware and software of the
stabilization system that transfers CHIME's reference clock to
a passive hydrogen maser. Section \ref{sec:validation} shows
the results of the suite of tests that validate the clock stabilization
system with VLBI-style observations between CHIME and the CHIME Pathfinder
\citep[an early small-scale prototype of CHIME recently 
outfitted as an outrigger test-bed, ][]{PathfinderOverview,leung2020synoptic}. 
Section~\ref{sec:no_maser_outrigger} presents an alternate clock solution
for outrigger stations that do not have the infrastructure to support
a hydrogen maser. Section~\ref{sec:conclusions} presents the conclusions.

\section{Instrument overview} 
\label{sec:inst_overview}

A detailed description of the CHIME instrument and the 
CHIME/FRB project is presented in
\cite{chime_overview_2021} and \cite{FRBSystemOverview}. In this section
we give a brief introduction to these systems focused on the
features that are relevant for FRB VLBI. We also
give an overview of CHIME/FRB Outriggers.

\subsection{CHIME and CHIME/FRB}
\label{subsec:chime_chimefrb}

CHIME is a hybrid cylindrical transit interferometer located at the 
Dominion Radio Astrophysical Observatory (DRAO) near Penticton, B.C., 
Canada. It consists of four 20 m $\times$ 100 m cylindrical reflectors 
oriented north-south and instrumented with a total of 1024 
dual-polarization feeds and low-noise receivers operating in the 
400-800~MHz band. The cylinders are fixed with no moving parts, 
so CHIME operates as a drift-scan instrument that surveys the northern 
half of the sky every day with an instantaneous field of view of
$\sim 120^\circ$ north-south by $2.5^\circ-1.3^\circ$ east-west.

Although CHIME's design was driven by its primary scientific goal to
probe the nature 
of dark energy by mapping the large-scale structure of neutral hydrogen in the universe across the redshift range $0.8\le z\le 2.5$,
its combination of high sensitivity and large field of view also
make it an excellent instrument to study the radio transient sky. Thus,
in its final stages of commissioning, the CHIME correlator was upgraded
with additional hardware and software backends
to perform additional real-time data 
processing operations for pulsar timing and FRB science.

The correlator 
\citep{2016JAI.....541005B,2016JAI.....541004B,2020JAI.....950014D}
is an FX design (temporal Fourier transform before spatial 
cross-multiplication of data), where the F-engine digitizes the 2048
analog inputs at 800~MSPS 
and separates the 400~MHz input bandwidth into 1024 
frequency channels with 390~kHz spectral resolution. 
The F-engine also implements the corner-turn network
that re-arranges the complex-valued channelized data (also known
as ``baseband'') before sending it to the X-engine that
computes a variety of data products for the different
real-time scientific backends: interferometric visibilities for the 
hydrogen intensity mapping backend \citep{chime_overview_2021}, 
dual-polarization tracking
voltage beams for the pulsar monitoring backend \citep{2020arXiv200805681C},
and high-frequency resolution power beams 
for the 21-cm absorption systems
backend \citep{2014PhRvL.113d1303Y} and for the CHIME/FRB
backend that triggers on highly dispersed radio transients 
to search for FRBs in real time \citep{FRBSystemOverview}. 
Additionally, a $\sim$36~s long memory buffer in the X-engine 
stores baseband data (2.56~$\mu$s time resolution, 
390~kHz spectral resolution, and 4-bit real~+~4-bit imaginary 
bit depth for the 2048 correlator inputs)
that can be saved to disk when the CHIME/FRB search pipeline detects
an FRB candidate, enabling polarization and high-time resolution
analysis of FRB events, as well as sub-arcminute localization 
precision \citep{michilli2020analysis}.
Eventually it will also enable VLBI localization
with CHIME/FRB Outriggers.

\subsection{CHIME/FRB Outriggers}
\label{subsec:chimefrb_outriggers}

The scientific goal of CHIME/FRB Outriggers is to provide 50~mas 
localization for nearly all CHIME detected FRBs
with sub-hour latency. This 
angular resolution is sufficient to determine galaxy hosts and
source environments, and is well matched to current best optical
follow-up observations. To this end, the CHIME/FRB collaboration
is currently building outrigger telescopes at distances ranging
from hundreds to several thousands of kilometers from DRAO.
The outriggers will be small-scale versions of CHIME, each with
about one eighth of CHIME's collecting area, the same field of view,
and tilted such that they monitor the same region of the sky as CHIME.

In contrast to traditional VLBI that is typically performed for known
targets with small fields of view and manageable data rates, 
the random nature of most FRBs requires the real-time processing 
of massive data rates in order to detect and localize these 
events in blind searches with wide fields of view. 
The baseband data rate of CHIME is
6.6~Tbit/s while that of each outrigger station will be 
0.8~Tbit/s. Since such high data rates cannot be continuously saved, 
the outriggers will adopt the triggered FRB VLBI approach demonstrated
in \cite{leung2020synoptic}, where each station buffers its local baseband 
data in memory and only writes it to disk upon receipt of a trigger 
from the CHIME/FRB real-time search pipeline
over internet links. The local data of each 
station is then transmitted to a central 
facility where the signals are correlated
together such that the outriggers operate with CHIME as an interferometric
instrument with the angular resolution of a telescope with an 
aperture of thousands of kilometers.

\section{Clock stability requirements} 
\label{sec:requirements}

Accurate timing is critical for VLBI since the localization of 
radio sources is ultimately derived from the relative time 
of arrival of signals at the telescope stations.
By synthesizing the available frequency channels it is
possible to obtain a statistical precision on the measured delay given
by \citep{1970RaSc....5.1239R}

\begin{equation}
    \label{eq:sigma_tau}
    \sigma_\tau^{\text{stat}} = \frac{1}{2\pi \cdot \text{SNR}\cdot \text{BW}_{\text{eff}}}
\end{equation}

\noindent where $\text{SNR}$ is the signal-to-noise ratio of the
VLBI event and $\text{BW}_{\text{eff}}$ is the effective bandwidth.
For the CHIME/FRB detection threshold\footnote{The SNR in VLBI
is related to the SNR at CHIME as 
$\text{SNR}/\text{SNR}_{\text{CH}} = \sqrt{2A_{\text{O}}/A_{\text{CH}}} = 1/2$ 
where $A_{\text{CH}}$ and $A_{\text{O}}$ 
are the collecting areas of CHIME and the
outrigger respectively, and the factor of $\sqrt{2}$ comes
from the difference in the detailed noise statistics of a 
cross-correlation compared to an auto-correlation 
\citep{2015A&C....12..181M}. 
While the CHIME/FRB real-time detection pipeline
has a detection threshold of $\sim 10$ \citep{michilli2020analysis}, 
the SNR rises by $\sim 50\%$
through the more detailed analysis of the saved baseband data. 
As such, we take the floor on the CHIME detection SNR to be 
$\text{SNR}_{\text{CH}} = 15$.} and bandwidth (\text{BW}) 
this corresponds to

\begin{equation}
    \label{eq:sigma_tau_1}
    \sigma_\tau^{\text{stat}} \approx \frac{184\text{ ps}}
    {\displaystyle \left(\frac{\text{SNR}}{7.5}\right)\left(\frac{\text{BW}}{400\text{ MHz}}\right)}.
\end{equation}

For a VLBI baseline $b$, a delay precision $\sigma_\tau$ corresponds to 
a statistical localization uncertainty

\begin{equation}
    \label{eqn:delay2pos_err}
    \sigma_\theta \approx \frac{c}{b} \sigma_\tau
\end{equation}

\noindent which gives $\sigma_\theta^{\text{stat}} \lesssim 11$~mas 
for a 1000~km baseline. However,
the (relative) delay measured by the interferometer 
includes not only the geometric delay
(which ultimately provides the source localization)
but also additional contributions that need to be accounted for such as 
propagation though the troposphere and ionosphere, baseline
errors, drift between clocks of different stations (clock timing errors),
and other instrumental delays. In practice,
the localization uncertainty of CHIME/FRB Outriggers
will be limited
by systematic errors due to uncompensated delay 
contributions, 
and particularly by errors in the 
determination of the dispersive delay due to the ionosphere.
Although the large observation bandwidth of the instrument 
helps to mitigate this effect, it still represents the 
most important challenge for
system stability at CHIME frequencies.

Our simulations indicate that we can reliably localize
FRB events to 50~mas which, for $b = 1000$ km,
corresponds to a delay error budget of $\sigma_\tau \approx 800$~ps.
Anticipating that the ionosphere
will be the main contributor to delay errors, the
clock timing error specification\footnote{This is not a 
hard upper limit, 
but rather a reasonable reference value that represents 
our goal to keep the clock timing errors well below the 800~ps 
total timing error budget.} has been set 
to $\sigma_\tau^{\text{clk}}\lesssim 200$~ps.

Note that for blind FRB searches, this
specification must be met at all times. Indeed,
it may not always be possible to find
a calibrator immediately after the 
detection of an FRB for phase referencing.
An additional complication is that 
stationary telescopes like CHIME and the outriggers
observe the sky as it transits through their field of view 
and thus cannot slew towards favorable calibrators. Therefore, it is especially important to have 
a reference clock that is reliable on the timescales required to 
connect an FRB detection to a calibrator observation, 
potentially hours later.

Although the list of steady radio sources 
potentially suitable for calibrating
low-frequency VLBI arrays with $\gtrsim 1000$~km baselines 
has significantly increased thanks to the 
ongoing LOw Frequency ARray (LOFAR) Long-Baseline Calibrator Survey 
\citep[LBCS, ][]{2015A&A...574A..73M,jackson2016lbcs}, 
during its initial stages CHIME/FRB Outriggers
will adopt a more conservative strategy 
relying mainly on bright pulsars for calibration.
Pulsars are compact, can be separated from the
steady radio background in the time domain,
and are sufficiently
abundant to be used as the primary sources for phase referencing.
Accordingly, the backend of each outrigger will also have the ability to 
form tracking baseband beams for pulsar analysis and calibration.
Recently, \cite{2021arXiv210705659C} demonstrated 
the potential of pulsars as calibrators 
for CHIME/FRB Outriggers through the triggered VLBI
detection of an FRB over a $\sim 3000$~km
baseline between CHIME and the 
Algonquin Radio Observatory (ARO) 10-m Telescope using
PSR B0531+21 in the Crab nebula for phase referencing.

We estimate that an FRB detection can be phase referenced
to an $\text{SNR}\gtrsim 15$ pulsar\footnote{This represents the
VLBI SNR after coherent addition of the pulses within
a single pulsar transit.}
within less than $\sim 10^3$~s. 
This timescale and the
relative clock timing error specification set the clock stability
(Allan deviation) requirement to

\begin{equation}
\label{eq:spec}
    \sigma_y(10^3~\text{s}) \lesssim 2 \cdot 10^{-13},
\end{equation}


As explained in Appendix~\ref{app:adev}, the Allan deviation
is a measure of stability commonly used in 
precision clocks and oscillators, with 
$\Delta t \cdot \sigma_y(\Delta t)$ roughly 
representing the rms of clock timing errors 
a time $\Delta t$ after calibration.
The specification in Equation~\ref{eq:spec} is typically obtained 
with hydrogen masers.
However, in Section~\ref{sec:no_maser_outrigger} we show
that even frequency references that initially do not meet this
requirement can be used as reference clocks for the outriggers
by interpolating timing solutions between calibrators.

\section{clock stabilization system} 
\label{sec:clock_stabilization}

In this section we discuss the considerations that led to
the current clock stabilization solution for CHIME, 
as well as the hardware and data analysis 
for the maser signal.

\subsection{Hardware/Software considerations}
\label{subsec:hw_sw_considerations}

The CHIME F-engine is implemented using the ICE
hardware, firmware and software framework \citep{2016JAI.....541005B}.
It consists of field programmable gate array (FPGA)-based motherboards
specialized to perform the data acquisition and channelization
of the 2048 CHIME analog inputs. The ICE motherboards are packaged in
eight crates with custom backplanes that implement the networking engine
that re-organizes and sends the baseband data to a dedicated 
graphics processing unit (GPU) cluster that performs the X-engine
operations. The outriggers will also use an ICE-based F-engine.

The data acquisition and signal processing of the 
F-engine are driven by a 
single 10~MHz clock signal provided by a
Spectrum Instruments TM-4D
global positioning system (GPS)-disciplined 
ovenized crystal oscillator. The GPS module also generates
an inter-range instrumentation group (IRIG)-B timecode signal
internally synchronized to the clock and 
that is used by the 
correlator to time stamp the data.
A low-jitter distribution system sends the clock and 
time signals to each ICE backplane and motherboard
and is ultimately used to generate the 
analog-to-digital converter (ADC) sampling clocks.
The time stamping process is implicit: The F-engine uses the
IRIG-B signal to synchronize the start of the data acquisition to
an integer second (up to the 10~ns resolution of the IRIG-B decoder
in the FPGA firmware) 
and it also tags each data frame with a frame counter
value. The X-engine time stamps the data by calculating 
the offset from the start time
based on the frame counter value and assuming a fixed
2.56~$\mu$s baseband sampling time.

\begin{figure}[h!]
    \centering
    \includegraphics[scale=.35]{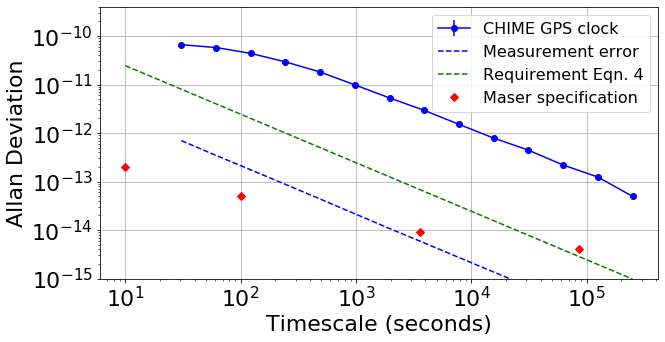}
    \caption{Allan deviation of the CHIME GPS clock
    and the DRAO hydrogen maser.
    Blue: Allan deviation of the CHIME GPS clock
    as measured with the clock stabilization system
    described in Section~\ref{sec:clock_stabilization}. A total of ten days
    of raw ADC data at 30~s cadence was collected for the
    measurement. Dashed blue: expected measurement 
    error contribution to the Allan deviation obtained from
    simulations of uncorrelated
    but time-dependent errors in the range $\sim 4-20$~ps rms
    (the range observed in the measured delays).
    Dashed green: stability requirement from Equation~\ref{eq:spec}
    assuming white noise delay errors. 
    Red: manufacturer-specified Allan deviation
    of the DRAO maser. 
    The CHIME GPS clock does not meet the stability requirements for
    FRB VLBI, but the DRAO maser does (Equation~\ref{eq:spec}).}
   \label{fig:chime_clk_adev}
\end{figure}

Figure~\ref{fig:chime_clk_adev} shows the Allan deviation of the 
CHIME GPS clock (blue line) as measured with the clock stabilization 
system described in Sections~\ref{subsec:maser_hw} and 
\ref{subsec:pipeline}.
The GPS disciplined crystal oscillator, being locked 
to a GPS time reference determined by a vast network of
atomic clocks, will eventually surpass the stability performance
of a single hydrogen maser on very long timescales
($\Delta t \gtrsim 10^6$~s).
On intermediate and long timescales ($\Delta t \sim 10^3-10^5$~s),
including the ones of interest for CHIME/FRB Outriggers,
the CHIME clock stability is dominated by
white delay
noise ($\sigma_y(\Delta t) \propto 1/\Delta t$) corresponding to
$\sim 6$~ns root mean square (rms) timing errors.
While the coherence of this frequency standard 
is sufficient for CHIME's operations
as a connected interferometer and for all its backends, 
the high precision needed for FRB VLBI 
requires the development of a more stable clock system.

As a continuously tracking global navigation satellite system (GNSS)
station and as part of the 
Western Canada Deformation Array \cite[WCDA, ][]{wcda1996_chen} and 
the Canadian Active Control System \cite[CACS\footnote{\href{http://cgrsc.ca/resources/geodetic-control-networks/canadian-active-control-system-cacs/}
{http://cgrsc.ca/resources/geodetic-control-networks/canadian-active-control-system-cacs}}, ][]{cacs1996_duval},
DRAO is equipped with an atomic frequency standard consisting
of a T4Science pH Maser 1008 passive hydrogen maser owned
and operated by Natural Resources Canada (NRCan).
The maser is installed in a seismic vault at the DRAO site
and has a primary output of 5~MHz (sine wave). It
is directly connected to a low noise distribution amplifier in the same
rack that serves as electrical isolation
for the maser and also derives multiple
copies of the 10~MHz reference signal (sine wave). 
NRCan has approved the use of
two of those signals for CHIME-related operations.

The manufacturer-specified Allan deviation of the DRAO maser is
shown in Figure~\ref{fig:chime_clk_adev} (red points).
The maser clearly exceeds the stability requirements for
FRB VLBI with CHIME (Equation~\ref{eq:spec}). 
Some of the outrigger sites will also have access
to hydrogen maser frequency references with similar performance.

Although in principle the ICE system can be operated with
an IRIG-B time signal that is not phase-locked to the 
10 MHz reference clock (such as an independent maser), 
an important restriction
to using the maser as the master clock for the 
CHIME correlator is the fact that both
the F-engine and X-engine software and the scientific data analysis
pipelines were developed on the premise that the clock and IRIG-B
signals are synced (e.g., to time stamp the data), 
something that cannot be guaranteed if the two
signals are generated by independent systems (the maser and the 
GPS receiver respectively). Even if the relative drift between clock and
time signals could be tracked, both the correlator and 
data analysis pipelines
would need to be updated to implement this change.
As such, we had to develop a clock
stabilization system that did not impact 
the normal operations of CHIME, its existing real-time backends,
and the other scientific teams.

\subsection{Maser signal conditioning}
\label{subsec:maser_hw}

The clock stabilization system designed for FRB VLBI with CHIME
keeps the current GPS-disciplined crystal oscillator as the
master clock and instead feeds the 
maser signal to one of the ICE ADC 
daughter boards so it is digitized by the crystal-oscillator-driven 
F-engine. The data is processed to
monitor the variations in the phase of the sampled maser signal
which correspond to variations in the relative delay
between the maser and the master clock.
By using this information to correct 
phase variations in the baseband data recorded at the time
of an FRB detection, the system effectively transfers the  
reference clock from the GPS disciplined oscillator to the more
stable maser signal during FRB observations.
As shown in Figure~\ref{fig:chime_clk_adev} (dashed blue line),
the noise penalty 
associated with the clock transfer operation
is essentially white 
($\sigma_y(\Delta t) \propto 1/\Delta t$)
on the timescales relevant for FRB VLBI 
and small ($\lesssim 20$~ps) compared
to the total clock timing error budget.

\begin{figure}[h!]
    \centering
    \includegraphics[scale=.39]{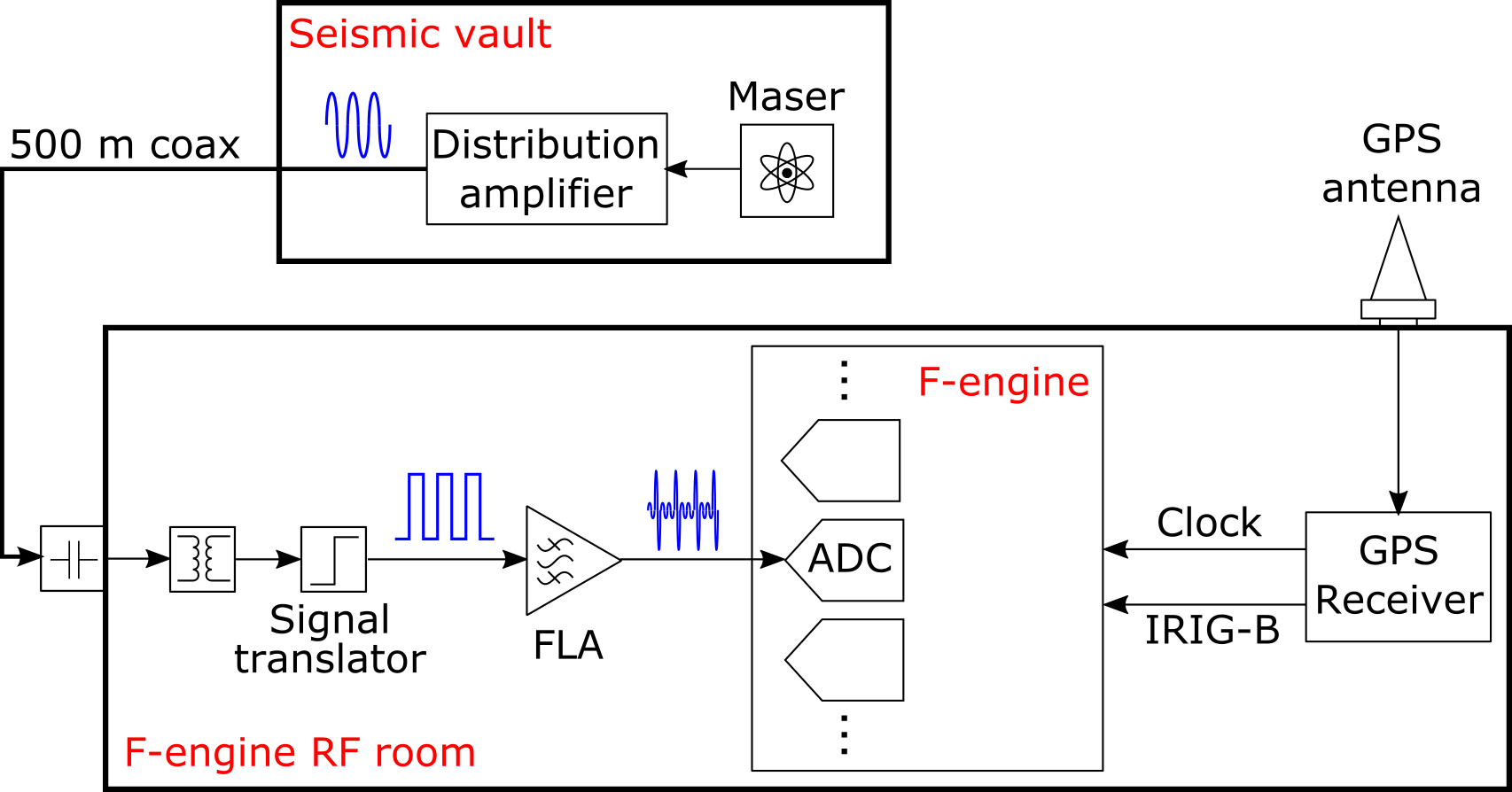}
    \caption{Maser signal path. The 10~MHz maser signal is transported 
    through $\sim 500$~m of buried coaxial cable from the seismic vault
    to one of the CHIME F-engine RF huts. There, the maser
    signal is conditioned to a waveform that can be
    digitized by the CHIME F-engine
    (see Section~\ref{subsec:maser_hw} 
    for details).}
    \label{fig:maser_setup}
\end{figure}

A block diagram of the maser signal path is shown in 
Figure~\ref{fig:maser_setup}.
The first point of access to the 10~MHz maser signal is the
low noise distribution amplifier within the seismic vault.
From there, the signal is transported through $\sim 500$~m of
buried coaxial cable to one of the two radio-frequency (RF)-shielded
huts that house the CHIME F-engine. 
We use the same type (LMR-400)
of low-loss coaxial cable used in the CHIME 
analog receivers and whose thermal susceptibility has been
extensively tested in the field. At the RF hut, the cable
interfaces with a ground block and the signal is then
carried inside the RF room using standard SMA cables 
where it is connected to an isolation transformer to refer 
the next stages to the F-engine crate ground.
One complication in the digitization of the maser signal is that the
ADC daughter boards that specialize the ICE system for
CHIME have a bandpass transfer function that strongly attenuates
signals below $\sim 100$~MHz. For this reason, instead of feeding
the maser signal directly to an ADC daughter board, the signal is
used to drive a low-noise sine-to-square wave signal translator
that generates 10~MHz harmonics well into the CHIME band. 
The output of the translator is then filtered
to the CHIME band using the same band-defining filter amplifier (FLA)
used in the CHIME receivers 
\citep{PathfinderOverview,chime_overview_2021}. 
Finally, the FLA is connected 
directly to one of the correlator inputs
where it is digitized at 800~MSPS with an 8-bit ADC.

\subsection{Clock stabilization pipeline}
\label{subsec:pipeline}

The FPGA within each ICE motherboard processes 
the data from its digitizers using the custom CHIME F-engine firmware
\citep[for details, see][]{2016JAI.....541005B,2016JAI.....541004B}. Briefly,
the raw ADC data from each input is passed to the
frequency channelizer module as frames of 2048 8-bit samples.
The channelizer forms the baseband data by splitting the 400 MHz input
bandwidth into 1024 frequency channels, each truncated to
a 4-bit real + 4-bit imaginary complex number. Additionally,
a probe sub-module within the channelizer 
can be configured to periodically capture a subset of 
the raw ADC data that is separately saved and typically 
used in CHIME for system monitoring.

By default, the CHIME F-engine software pipeline 
saves one raw ADC frame (2.56 $\mu$s of data)
from each input every 30~s, but this cadence can 
be modified before starting a
data acquisition. 
The clock stabilization pipeline 
extracts the raw ADC frames
from the maser input, Fourier transforms each frame via a
Fast Fourier Transform (FFT), and separates the frequency channels
corresponding to the harmonics of the 10 MHz signal in the CHIME
band. 
The quality of each harmonic is assessed based on its 
signal-to-quantization-noise ratio (SQNR) and its 
susceptibility to spurious aliased harmonics (relevant for
harmonics near the edges of the CHIME band).
Low quality harmonics are discarded.

Since the ADC that digitizes the maser signal uses the GPS clock
as the reference for sampling, the variations in the delay of the
sampled maser signal represent the delay variations of the
GPS clock with respect to the maser, the latter of which is more
stable on short and intermediate timescales. These delay
variations $\Delta\tau(t)$ will induce 
phase variations $\Delta \phi (t, \nu)$ in the maser harmonics of
the form

\begin{equation}
    \label{eq:delay_to_phase_var}
    \Delta \phi (t, \nu) = 2\pi\nu \Delta\tau(t)
\end{equation}

\noindent where $\nu$ is the harmonic frequency.

Since we are interested in the GPS clock delay variations
relative to the delay at the time of VLBI calibration, 
the phase of the maser harmonics is initially referenced to
the phase of a frame close to calibration time. Then for each frame,
a line described by Equation~\ref{eq:delay_to_phase_var} is fit to
the phase as a function of harmonic frequency to recover the 
GPS clock delay (relative to the maser)
as a function of time. 
The dominant component of the 
recovered delay $\Delta\tau(t)$ is a slow 
linear drift as function of time ($\sim 50$~ns/day)
corresponding to a constant offset of the maser frequency from
10~MHz. This linear trend is removed from $\Delta\tau(t)$
since it is due to the maser frequency calibration and not due to the 
instability of the GPS clock.

\begin{figure}[h!]
    \centering
    \includegraphics[scale=.42]{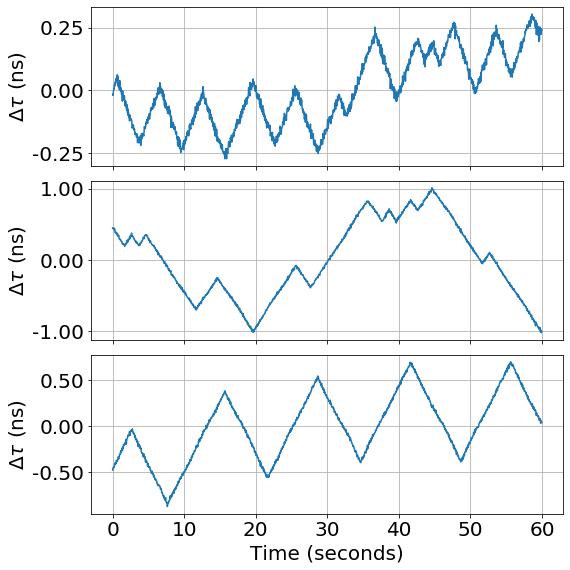}
    \caption{Three examples of the behavior of the
    GPS clock delay on timescales of a few seconds as measured by
    the clock stabilization system with respect to the DRAO maser.
    The raw ADC data cadence for this measurements was 40~ms.
    The measurement errors are in the range $\sim 2-13$~ps.
    The characteristic triangle wave pattern is due to
    the algorithm that disciplines the crystal oscillator
    in the GPS unit. 
    The algorithm works by counting the number of clock cycles between 
    successive GPS PPS pulses and adjusting
    the crystal's temperature to ensure ten million 
    counts between pulses.
    The size of the temperature tuning steps changes depending on
    the tuning history of the oscillator.}
    \label{fig:tuning_jitter}
\end{figure}

The captured raw ADC
frames are only a small fraction of the available CHIME data; thus,
the times at which GPS clock delay measurements are available are
not necessarily aligned with the times of a calibration observation
or an FRB detection. This means that the GPS clock delay timestream
must be interpolated in order to find the clock 
contribution to the total delay measured in a VLBI observation.
We use linear interpolation
to find the GPS clock delay at any arbitrary time, a method
that is motivated by the short timescale behavior of the
clock delay variations. 
Figure~\ref{fig:tuning_jitter} shows a 
few examples of the behavior of the
GPS clock delay on timescales of a few seconds as measured by
the clock stabilization system.
On these timescales the timing variations are dominated by
the tuning jitter generated by the algorithm that disciplines
the crystal oscillator. In essence, the algorithm works 
by counting the number of clock cycles between 
successive GPS receiver pulse-per-second (PPS) pulses and adjusting
the crystal's temperature to ensure ten million counts between pulses.
The size of the temperature tuning steps is progressively 
reduced as the crystal oscillator frequency approaches 10~MHz.
As shown in Figure~\ref{fig:tuning_jitter}, 
this discipline procedure gives a characteristic
triangle-wave shape to the tuning jitter, and although in a 
perfectly-tuned oscillator the transitions should occur every second,
in practice we observe that they can take longer.
Thus, as long as the GPS clock delay is sampled at cadences
below $\sim 500$~ms we can track the tuning jitter features and
a linear interpolation provides a good approximation to
the true delay at any time.

As the current version of the F-engine control software only allows 
saving raw ADC data for all the correlator inputs at the time, raw ADC
data at cadences below 10~s cannot be saved during normal
telescope operations and are restricted to operations during
times scheduled for
hardware maintenance and software upgrades.
A modification to the F-engine control software is ongoing to
allow saving fast-cadence raw ADC data for the maser input while
keeping the default cadence for the remaining correlator inputs,
a change that does not impact the the normal operations of the
correlator and the data-analysis pipelines.

It is also possible to process baseband
data directly to extract the maser signal and measure the GPS clock
delay variations. The operation is very similar
to that of raw ADC data, except that the maser data has already
been transformed to the frequency domain by the F-engine. 
Since in this case 
most maser harmonics do not lie exactly in the center of 
a frequency channel, the pipeline selects the closest F-engine frequency
channel. Then for each selected channel, it
performs an additional channelization by
using an FFT along the time domain to isolate the harmonic frequency.

Although working with baseband data is logistically convenient in 
cases where we need to test the performance of the 
clock stabilization system (see Section~\ref{sec:validation}),
during regular operations the current system is designed to work mostly
with raw ADC data. This is mainly because a baseband dump for an
FRB event is typically collected in 
$\sim 100$~ms segments at different
times for each frequency channel in order to account for the
dispersion delay of the transient, with a total event duration lasting
tens of seconds \citep{michilli2020analysis}. 
This leaves only a few megahertz of bandwidth
available at any particular instant, making the monitoring 
of clock delay variations more challenging. Furthermore,
when using baseband dumps we still need to rely on the 
continuously saved raw ADC data to track and correct the
long-timescale linear drift of the maser.

\section{Validation of the clock stabilization system} 
\label{sec:validation}

We tested the reliability of clock stabilization system
by installing it in the Pathfinder telescope and 
comparing maser-based measurements of the CHIME-Pathfinder 
relative clock drift to independent measurements
obtained from VLBI-style observations of steady
radio sources.

\subsection{The Pathfinder as an outrigger} 
\label{subsec:pathfinder}

The Pathfinder is presented in detail in \cite{PathfinderOverview}.
It is a small-scale prototype of CHIME with identical
design and field of view, and it has 
the same collecting area of the
outriggers under construction. 
The telescope is located $\sim 400$~m from CHIME
and was constructed
before CHIME as a test-bed for technology development.
With the same correlator architecture as CHIME, the Pathfinder
operates as an independent connected interferometer with 
its own GPS-disciplined clock.
Recently, \cite{leung2020synoptic} repurposed the Pathfinder
as an outrigger to demonstrate the feasibility of
triggered FRB VLBI for CHIME/FRB Outriggers.
The Pathfinder correlator is now equipped with a custom
baseband data recorder capable of processing 
one quarter of the CHIME band and programmed 
to write its local baseband data to disk 
upon receipt of a trigger from CHIME/FRB.
We also connected the additional copy of the maser signal from the
seismic vault to one of the Pathfinder correlator inputs using
a signal path identical 
to that of CHIME and shown
in Figure~\ref{fig:maser_setup}
(except for the transport cable
which is longer for the Pathfinder setup).

\subsection{Comparison to interferometric observations} 
\label{subsec:CygA}

The clock stabilization system measures the delay variations
of the CHIME and Pathfinder clocks by processing the
maser data from each telescope as described in 
Section~\ref{sec:clock_stabilization}. The delays
from each clock are then interpolated to the observation times
so the relative clock drift can be
tracked over time\footnote{If the maser data 
comes from simultaneous baseband dumps 
instead of raw ADC samples then clock
delays from the two telescopes can be directly compared 
without interpolation.}. 
An independent way to measure the CHIME-Pathfinder relative clock
delay is to interferometrically track a known point source over time  
using both telescopes and their independently-running backends. 
If we properly account for all the other contributions
to the measured interferometric delay
(geometric, ionosphere, etc.) as the source transits through the
field of view of the two telescopes, then any residual delay
should correspond to the relative drift between the two clocks. 
If the clock stabilization system is robust, 
its measurements should agree very 
closely with the interferometric measurement, 
which we use as a standard.

\subsubsection{Short timescale test} 
\label{subsubsec:short_timescale_test}

We used Cygnus-A (henceforth referred to as CygA)
for the VLBI-style observations since it
is the brightest radio source seen by CHIME that
is unresolved on a CHIME-Pathfinder baseline.
For the first test, we programmed the CHIME/FRB backend to 
trigger short baseband dumps 
simultaneously for CHIME and the Pathfinder
during a single CygA transit. In this way, we collected 
seven 10~ms-long baseband dumps, spaced by a minute, while 
the source was in the field of view.
The observation was carried out in November 2020 during a day
scheduled for instrument maintenance, so we were also able
to collect raw ADC maser data at 200~ms cadence with the two telescopes.

Since the telescopes are co-located, they experience a common ionosphere,
suppressing relative ionospheric fluctuations (we see no evidence for these
in our observations). Thus, the residual delay in the 
interferometric visibility after accounting for 
the geometric contribution gives a measurement of the relative
clock drift. The residual interferometric delays
are calculated using a procedure identical to 
that described in \cite{leung2020synoptic}.
In summary, each telescope is internally calibrated
(to measure the directional response of each antenna in the 
telescope array) using a separate observation of a bright point source 
\citep{chime_overview_2021}.
Then, for each telescope and baseband dump, the data is
coherently summed over antennas to form
a phased-array voltage beam towards CygA. 
The beamformed data from CHIME and the Pathfinder 
is then cross-correlated on a frequency-by-frequency basis
to form the complex visibility. The phase of each visibility is
compensated for the geometric delay. 
The variance of each visibility is found empirically by
splitting each baseband dump into short time segments,
computing the visibility for each segment, and calculating the
variance over segments as function of frequency.
The seven visibilities are phase-referenced to that of the first
baseband dump since we are only interested in
changes of the relative clock delay.
Finally, the  wideband fringe-fitting procedure to find residual delay
performs a least-mean-squares fit of a complex exponential
with linear phase and frequency-dependent amplitude
to the measured visibilities.

\begin{figure}[h!]
    \centering
    \includegraphics[scale=.35]{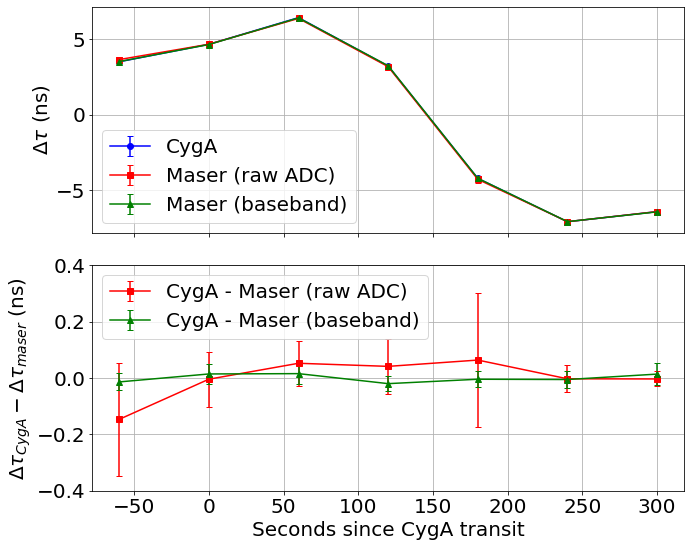}
    \caption{Comparison of the CHIME-Pathfinder relative clock delay 
    inferred via the clock stabilization system and interferometric
    observations from a single transit of CygA. 
    Top: relative clock delay (in ns) 
    as function of time as inferred from
    CygA baseband data (blue), 
    raw ADC maser data (red) and maser baseband data (green).
    Bottom: Difference between sky-based and maser-based measurements
    of the relative clock delay
    (raw ADC maser data in red, baseband maser data in green),
    demonstrating agreement between the two methods. The large error bars
    for all but the last point in the raw ADC data analysis (red)
    are due to current limitations of the Pathfinder raw ADC
    acquisition system (see Section~\ref{subsubsec:short_timescale_test} 
    for details). The error bar in the
    last red point of the top plot ($\sim 14$~ps) 
    is representative of the expected accuracy
    of the clock stabilization system using raw ADC data.}
   \label{fig:short_timescale}
\end{figure}

The top panel of Figure~\ref{fig:short_timescale} shows the resulting
comparison between the CHIME-Pathfinder relative clock delays
calculated from interferometric observations (blue) 
and those found through the clock stabilization system
(from raw ADC data in red, from baseband data in green). 
The $1\sigma$ error bars are too small to be visible in the plot but 
they are in the range $\sim 17-22$~ps for CygA measurements,
$\sim 14-238$~ps for raw ADC maser data measurements,
and $\sim 18-33$~ps for baseband maser data measurements.
Measurements with the clock stabilization system 
show excellent agreement with the sky. This is further
highlighted in the bottom panel of Figure~\ref{fig:short_timescale}
that shows the difference between sky-based and maser-based measurements
of the relative clock delay 
(raw ADC maser data in red, baseband maser data in green).

The large error bars for all but the last point in the 
raw ADC data analysis (red) 
are dominated by the error in the measurements of 
the maser delay at the Pathfinder. These can be traced back to current 
limitations of the Pathfinder raw data acquisition system, which 
occasionally drops packets when we collect raw ADC data at fast cadence
for the reasons explained in Section~\ref{subsec:pipeline}.
This limitation will be solved in the next upgrade of the
F-engine control software. For the observation times that
fell within sections of missing Pathfinder raw ADC data 
(the longest of which was $\sim 80$~s),
the delay values were obtained by performing 
a smoothing spline interpolation based on the available
measurements. To estimate the uncertainty in the delay values
obtained with this method we analyzed a segment of the
delay timestream for which there were no gaps due to dropped
packets, $\sim 10$~min before the sky observations.
The errors were found using a procedure similar to the one 
used in Section~\ref{sec:no_maser_outrigger} to evaluate
the performance of alternate reference clocks, 
where we introduce artificial gaps in the delay timestream and
analyze the statistics of an ensemble of interpolation
residuals. Only the raw ADC delays (red)
for the last observation time could be measured
using the default interpolation for both telescopes
(see Section~\ref{subsec:pipeline}). The
uncertainty for this measurement is $\sim 14$~ps, and
is representative of the expected accuracy
of the clock stabilization system using raw ADC data.

\subsubsection{Long timescale test} 
\label{subsubsec:long_timescale_test}

To test the performance of the clock stabilization system
on long timescales we collected five CygA baseband dumps, 
each 10~ms in duration, spaced one minute apart, 
for nine days in a row for a total of 45 delay measurements.
The observations were carried out in April 2021 during
normal CHIME operations so we relied on baseband
maser data for delay measurements with the clock stabilization
system for the reasons explained in Section~\ref{subsec:pipeline}.
Both interferometric and maser-based delays 
are calculated using the same procedure as in the 
short timescale test described in 
Section~\ref{subsubsec:short_timescale_test}, with the 
visibilities phase-referenced to one of the observations
on the fifth day.

\begin{figure}[h!]
    \centering
    \includegraphics[scale=.35]{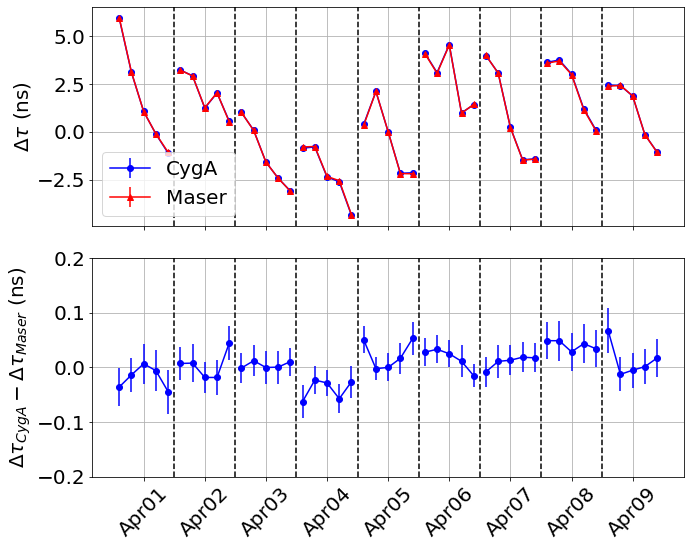}
    \caption{Top: Comparison of the CHIME-Pathfinder relative clock delay 
    inferred via the clock stabilization system (red) and interferometric
    observations (blue) of from multiple transits of CygA. 
    For each transit, we made five measurements of the
    relative clock delay, spaced by one minute,
    for nine days in a row.
    Bottom: Difference between sky-based and maser-based 
    measurements of the relative clock delay. 
    The two methods show excellent agreement
    on short (minute) and long (many days) timescales.
    This indicates that the clock stabilization system 
    we have implemented can track clock delay variations with 
    better than $\sim 30$~ps rms level precision.}
   \label{fig:long_timescale}   
\end{figure}

The results of the long timescale test are shown in Figure~\ref{fig:long_timescale}. The interferometric
measurements agree with the maser-based measurements 
at the $\sim 30$~ps rms level, demonstrating that 
after correction with the clock stabilization system 
the resulting reference clock is stable 
over timescales of more than a week, 
and that the signal
chain used to inject the maser signal into the correlator 
is not a limitation for the system's performance
in CHIME/FRB Outriggers.

\section{A reference clock for outrigger stations without a maser}
\label{sec:no_maser_outrigger}

The clock stabilization system allows us to inject external 
reference clock signals into radio telescopes which share the 
CHIME correlator architecture. 
In addition to its use in the new outriggers, 
the system enables reference clocks 
to be swapped out at existing telescopes like CHIME
and legacy systems like the Pathfinder 
without making major changes to the software framework or existing 
scientific backends, while expanding the telescopes' capabilities to 
include VLBI. 

Most outriggers will also have access to 
hydrogen maser frequency references that can be used in the
same way as CHIME (Section \ref{sec:clock_stabilization})
to meet the stability requirements
for FRB VLBI. However, we still 
need to address the possibility that certain outrigger 
stations may be built at locations (e.g., greenfield land) that will 
lack the infrastructure to support a hydrogen maser.
In this scenario, alternate reference signals 
(e.g., from rubidium microwave oscillators) 
can also be injected into the correlator to track and compensate 
for GPS clock drifts. The performance of these oscillators is
inferior to that of a hydrogen maser, but they are still
more stable than the CHIME GPS clock on the timescales relevant 
for FRB VLBI. They are also less expensive and
more readily available than a maser.
Even if these
frequency references can potentially
be used directly as the correlator master clock since they typically
come in units that can provide GPS disciplining as well as absolute time,
it is still desirable to use them separately as free-running clocks
for short and intermediate timescale observations. Not only are they inherently more stable than the primary CHIME clock, but they are not subject to short-timescale tuning jitter when not locked to GPS.

Equation~\ref{eq:spec}
provides a convenient way to determine whether an off-the-shelf 
clock meets the requirements for FRB VLBI
by simply reading the $\sigma_y(10^3~\text{s})$ 
value from the unit's datasheet.
Passive hydrogen masers meet and exceed this requirement.
However, this specification
was derived from Equation~\ref{eq:adev_wn_app}
which assumes uncorrelated clock timing errors and 
perfect calibration measurements. 
In practice, the actual clock timing errors
will depend on aspects not necessarily captured by
this equation including
the detailed statistics of the delay variations,
the methods used to estimated them, the
timing and accuracy of the calibration measurements,
and the technique that we use to inject the clock to our system.
These aspects become relevant when 
the frequency standard does not clearly exceed the
specification in Equation~\ref{eq:spec}.

As part of the implementation of the clock stabilization system,
\cite{timing_pipeline_scary_2021} developed a software package
with methods to
determine the suitability of precision clocks for
VLBI with transit telescopes like CHIME/FRB Outriggers. 
These methods take 
into account the details of the 
noise processes that determine the stability of the clocks
and simulate realistic timing
calibration scenarios.
The basic input to the software is a timestream that
represents the delay variations as function of 
time of the clock under test.
The delay timestream data can be either from measurements or from
simulations; in the latter case the software provides tools to
generate timestreams described by
combinations of power-law noise processes
commonly observed in precision clocks and oscillators including
white phase modulation noise, white frequency modulation noise, 
flicker frequency modulation noise, 
and random walk frequency modulation noise
\citep{1539968}.
Similarly, the software provides tools to generate
delay timestreams from a set of Allan deviation measurements,
which is convenient to evaluate the performance of a 
clock based on its manufacturer specifications.
In this case, it is assumed that the delay variations
are described by a combination of power-law noise processes
where the weight of each noise component is found by
fitting the Allan variance data
to a model consisting of a linear combination of the 
Allan variances of the previously described noise processes.

A calibrator is parametrized
by its observing time, number of
clock timing measurements, and $\text{SNR}$ per transit. 
For example, for a calibrator at the equator
the observing time with CHIME
is $\sim 6$~min, and with a $\sim 2$~min
integration time we would have three 
delay measurements per transit. The $\text{SNR}$ determines
the uncertainty in the calibration delay measurements
(see Equation~\ref{eq:sigma_tau}).
Given a timescale $\Delta t_{cal}$ that represents the 
maximum expected time
separation between calibrators, the method masks a
random $\Delta t_{cal}$-long section 
of the delay timestream, interpolates
using a best-fit function determined from 
the available calibration measurements at each end
of the masked section, and keeps the interpolation residuals.
The process is repeated a 
configurable number of times
to obtain a statistical ensemble of interpolation residual
timestreams, each of length\footnote{In
practice there is an additional overhead equivalent to one integration.} 
$\sim \Delta t_{cal}$. 
As the default metric of the
stability of the clock at $\Delta t_{cal}$ timescales, the method
uses the largest value of the ensemble standard deviation
in the interval $\left [0, \Delta t_{cal} \right]$.
Other metrics of performance are available.
Since throughout the paper
we have used the convention that $\Delta t$ represents the
time between a calibration and an observation, this metric
can be interpreted as an
estimate of the largest clock timing rms error 
for $\Delta t$ up to $\sim \Delta t_{cal}/2$.

Different interpolation methods are available including
linear fit, smoothing spline, and 
nearest available calibrator.
The fitting weights are determined
by the calibrator $\text{SNR}$ and the level of 
the noise added by the timing stabilization system.

\begin{figure}[h!]
    \centering
    \includegraphics[scale=.35]{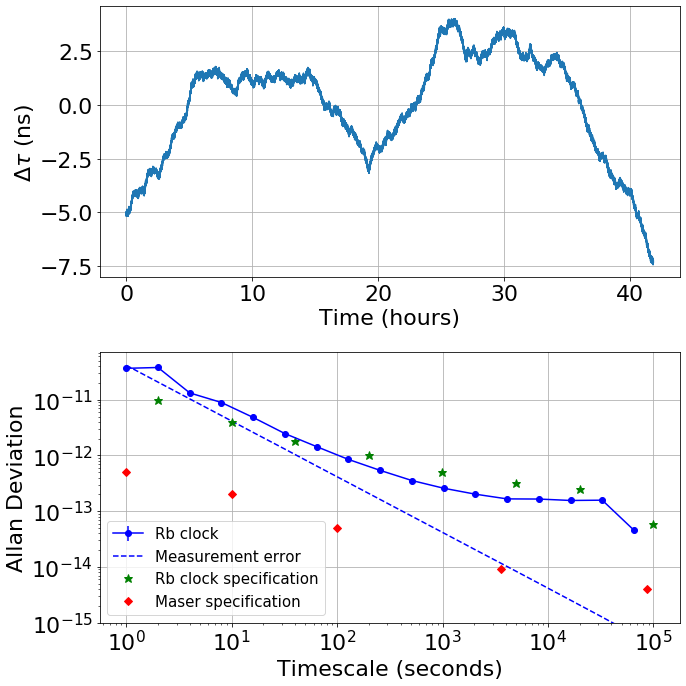}
    \caption{Top: Measured delay variations of the
    rubidium oscillator tested as a candidate reference clock
    for outriggers without a maser. 
    Bottom: Measured Allan deviation of the Rb clock (blue),
    measurement error (dashed blue), and
    manufacturer-specified Allan deviation of the 
    Rb clock (green points) and the DRAO maser (red points).
    The measured Allan deviation of the Rb clock is consistent
    with the specification at intermediate and long timescales.
    At short timescales the noise of the timing stabilization system
    dominates the performance, but it is still small ($\sim 10-30$~ps)
    compared to the clock timing error budget. This confirms that
    the hardware of the clock stabilization system
    is not a limitation for clock performance in
    CHIME/FRB Outriggers.}
   \label{fig:merid_clk_adev}
\end{figure}

As a candidate for outriggers without a maser,  we evaluated the
performance of the EndRun Technologies Meridian II US-Rb 
rubidium oscillator. The unit was installed 
in the Pathfinder RF room and connected to a
separate input of the correlator so we could test its 
performance against the DRAO maser under conditions
comparable to those of a typical outrigger. A signal
conditioning chain identical to 
the maser was used  (signal translator + FLA). 
We collected $\sim 42$~h of raw ADC data
at 1~s cadence for both the maser and the Rb clock
and used the clock stabilization pipeline to extract
the clock delay variations relative to the maser.
The top panel of Figure~\ref{fig:merid_clk_adev} 
shows the measured delay variations after
removing slow linear drift component due to the maser
(see Section~\ref{subsec:pipeline}). 
The bottom panel
shows the corresponding Allan deviation of the Rb clock (blue),
the measurement error (dashed blue), and
the manufacturer-specified Allan deviation of the 
Rb clock (green points) and the DRAO maser (red points).
For $\Delta t \lesssim 40$~s the 
variations are dominated by
the noise associated to the timing stabilization system,
which is in the range $\sim 10-30$~ps, small compared
to the $\sim 200$~ps clock timing error budget.
At longer timescales the measured 
Allan deviation of the Rb oscillator is consistent
with the manufacturer specification. These
results are also consistent with direct
Allan deviation measurements
of the Rb oscillator performed with a phase noise analyzer,
confirming that
the hardware of the stabilization system
is not a limitation for clock performance in
CHIME/FRB Outriggers.

Note that if we rely only on the manufacturer-specified 
Allan deviation, the Rb clock does not meet the
requirement in Equation~\ref{eq:spec}. This justifies
a more detailed analysis of the clock performance to 
determine whether it can still be used as a frequency
reference for the outriggers without a maser.

\begin{figure}[h!]
    \centering
    \includegraphics[scale=.36]{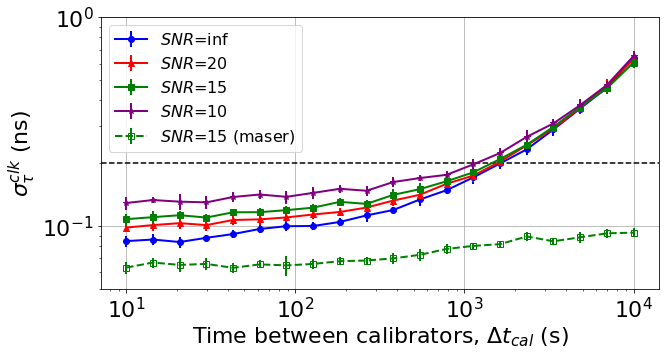}
    \caption{Projected clock errors, $\sigma_\tau^{\text{clk}}$, of the
    Rb clock as function of the time between calibrators
    $\Delta t_{cal}$ from measured delay variations and 
    simulations of
    realistic  timing  calibration  scenarios.
    This metric represents an estimate of the 
    largest clock timing error for 
    $\Delta t$ up to $\sim \Delta t_{cal}/2$ 
    (see Section~\ref{sec:no_maser_outrigger} for details).
    The dashed black horizontal line represents 
    $\sigma_\tau^{\text{clk}} = 200$~ps.
    Even in the most conservative scenario 
    where we assume that all the 
    calibrators have $\text{SNR} = 15$ (solid green), 
    the Rb clock timing errors stay below
    $225$~ps up to $\Delta t \sim 10^3$~s
    by interpolating between timing solutions,
    meeting the requirements for FRB VLBI with CHIME/FRB Outriggers.
    }
    \label{fig:meridian}
\end{figure}

The measured delay timestream was analyzed with the
software package described above and in \cite{timing_pipeline_scary_2021}
to evaluate the expected performance
of the Rb clock under different calibration conditions.
The results are shown in Figure~\ref{fig:meridian}.
For the analysis we assumed that the calibrator observing time
was 9~min (roughly the value at CHIME's zenith) 
with two integrations per observation.
The best performance is obtained with 
linear interpolation between calibrators.
We tested calibrator $\text{SNRs}$ of 10 (purple), 
15 (green), 20 (red), and $\infty$ (blue), 
the latter representing the case where the clock performance
is not limited by calibration errors. 
For comparison we also show the projected clock timing errors
for the DRAO maser using synthetic data generated from
the manufacturer-specified 
Allan deviation with $\text{SNR}=15$ (dashed green).
Figure~\ref{fig:meridian} shows that, 
even in the most conservative scenario
where we assume that all the calibrators have $\text{SNR} = 15$, 
the Rb clock timing errors stay below
$225$~ps up to $\Delta t \sim 10^3$~s
by interpolating between timing solutions (solid green).
This clock timing error is still well below
the total timing budget of $\sim 800$~ps,
leaving enough room to handle ionospheric delay errors.

\section{Conclusions}
\label{sec:conclusions}

We developed a clock stabilization system for CHIME/FRB Outriggers that 
allows synchronization of CHIME and outrigger 
stations at the $\sim 200$~ps level 
on short and long timescales. This meets the requirements 
for 50~mas localization of FRBs detected with the 
CHIME/FRB real-time pipeline. Our proof-of-principle clock 
transfer has demonstrated that a variety of different data 
products can be used for precise time transfer from an external 
reference clocks into data acquisition backends using the ICE 
framework. This method is minimally invasive to existing 
telescopes like CHIME and the Pathfinder,
expanding the capabilities of these instruments to include VLBI
without impacting their existing scientific backends.
It also allows for 
increased flexibility and modularity for future systems such 
as those at CHIME outriggers.
For outriggers that do not have the infrastructure to
support a hydrogen maser, we demonstrated that
it is still possible to meet the required clock stability 
specification by using alternate reference clocks
and interpolating timing solutions between calibrations.

\begin{acknowledgments}

We acknowledge that CHIME is located on the traditional, 
ancestral, and unceded territory of the Syilx/Okanagan people.

We are grateful to the staff of the Dominion Radio
Astrophysical Observatory, which is
operated by the National
Research Council Canada. 
CHIME is funded by a grant from the Canada Foundation 
for Innovation (CFI) 2012 Leading Edge Fund (Project 31170) 
and by contributions from the provinces of British Columbia, 
Qu\'ebec and Ontario. The CHIME/FRB Project is funded by a 
grant from the CFI 2015 Innovation Fund (Project 33213) and 
by contributions from the provinces of British Columbia and 
Qu\'ebec, and by the Dunlap Institute for Astronomy and 
Astrophysics at the University of Toronto. 
Additional support was provided by the Canadian 
Institute for Advanced Research (CIFAR), McGill 
University and the McGill Space Institute via the 
Trottier Family Foundation, and the University of 
British Columbia. 
The CHIME/FRB Outriggers program is funded by 
the Gordon and Betty Moore Foundation
and by a National Science Foundation (NSF) grant (2008031).
FRB research at MIT is supported by an NSF grant (2008031).
FRB research at WVU is supported by an NSF grant (2006548, 2018490).
J.M.P. is a Kavli Fellow.
C.L. was supported by the U.S. Department of Defense (DoD) 
through the National Defense Science \& Engineering Graduate 
Fellowship (NDSEG) Program.
V.M.K. holds the Lorne Trottier Chair in Astrophysics \& 
Cosmology and a Distinguished James McGill Professorship 
and receives support from an NSERC Discovery Grant and 
Herzberg Award, from an R. Howard Webster Foundation 
Fellowship from CIFAR, and from the FRQNT Centre de 
Recherche en Astrophysique du Qu\'{e}bec.
J.L.S. acknowledges support from the Canada 150 programme.
\end{acknowledgments}

%

\vspace{5mm}
\facilities{CHIME}


\software{
\texttt{AllanTools}~\citep{2018ascl.soft04021W}}




\appendix
\section{Clock stability and Allan Deviation}
\label{app:adev}

On timescales of a second and larger, clock stabilities are 
usually quoted in terms of the Allan variance, or its 
square-root the Allan deviation \citep{1446564}. 
In this appendix, 
we show how the clock stability required for FRB VLBI 
relates to the Allan deviation.

Let the ``true'' time be $t$, and the difference in seconds between 
what our clock reads and the true time be $x(t)$.  
In general, we will not know what the 
clock timing error $x(t)$ is for all time, 
but for CHIME/FRB Outriggers we will measure it whenever 
we do an on-sky calibration.  
If we calibrate at two points in time, we naturally would 
like to predict the timing error at the midpoint between 
the two calibrations. In particular, if
the calibration times are $t$ and $t+2\cdot \Delta t$, 
then we want to estimate $x(t+\Delta t)$ 
given $x(t)$ and $x(t+2\cdot \Delta t)$.  
The simplest assumption that we can make is that our 
clock runs at a constant rate between $t$ and $t+2\cdot \Delta t$, 
in which case our prediction for $x(t+\Delta t)$ will 
be $\left [ x(t)+x(t+2\cdot \Delta t) \right ]/2$. 
The variance of the timing error
estimate halfway between the calibrations will then be:

\begin{equation}
\label{eq:var_tau_app}
\left < \left [x(t+\Delta t)-\frac{x(t)+x(t+2\cdot \Delta t)}{2} \right ]^2 \right >=\frac{1}{4} \left < \left [x(t) -2x(t+\Delta t)+x(t+2 \cdot \Delta t) \right ]^2 \right >.
\end{equation}

The Allan variance is defined to be 

\begin{equation}
\label{eq:adev_app}
\sigma_y^2 (\Delta t) \equiv \frac{1}{2 \cdot \Delta t^2}\left < \left [x(t)-2x(t+\Delta t)+x(t+2 \cdot \Delta t) \right ]^2 \right >,
\end{equation}

\noindent so we can now relate our timing error 
variance halfway between the calibrations directly to the Allan variance:

\begin{equation}
\label{eq:var_tau_app1}
\left < \left [x(t+\Delta t)-\frac{x(t)+x(t+2\cdot \Delta t)}{2} \right ]^2 \right >=\frac{\Delta t^2}{2}\sigma_y^2(\Delta t).
\end{equation}

In general, and up to factors of order unity, the variance of the 
timing error will be determined by the product 
$\Delta t^2 \cdot \sigma_y^2(\Delta t)$, 
so

\begin{equation}
\label{eq:sigma_tau_app}
\sigma_x(\Delta t)\approx \Delta t \cdot \sigma_y(\Delta t).
\end{equation}

\noindent is typically used as an approximation to the
standard deviation of the clock timing error after a time 
$\Delta t$ \citep{5570702,4314804,6312409}.
In practice, the actual errors will
depend on the details of the prediction algorithm and the 
noise processes that dominate the 
stability of the clock.
For the particular case of white timing noise it follows from
Equation~\ref{eq:adev_app} that

\begin{equation}
\label{eq:adev_wn_app}
\sigma_y^{WN}(\Delta t) = \sqrt{3} \frac{\sigma_x}{\Delta t}.
\end{equation}

From Equation~\ref{eq:adev_app} the Allan
deviation is dimensionless, so it is telling us the 
fractional uncertainty in our clock between calibrations. 
Quantitatively, if our pulsar calibrations are separated by 
$2\cdot \Delta t=2000$~s, 
and the limit on the 
\emph{differential} timing residual at the midpoint is $200$~ps rms
(Section~\ref{sec:requirements}), then our Allan deviation requirement is 

\begin{equation}
\label{eq:adev_example_app}
\sigma_y(10^3~\text{s})\approx \sqrt{\frac{3}{2}}\cdot \frac{2\cdot 10^{-10}~\text{s}}{10^3~\text{s}}\approx 2 \cdot 10^{-13}.
\end{equation}

\bibliography{references}{}

\newcommand{\noop}[1]{}
\begin{thebibliography}{}
\expandafter\ifx\csname natexlab\endcsname\relax\def\natexlab#1{#1}\fi
\providecommand{\url}[1]{\href{#1}{#1}}
\providecommand{\dodoi}[1]{doi:~\href{http://doi.org/#1}{\nolinkurl{#1}}}
\providecommand{\doeprint}[1]{\href{http://ascl.net/#1}{\nolinkurl{http://ascl.net/#1}}}
\providecommand{\doarXiv}[1]{\href{https://arxiv.org/abs/#1}{\nolinkurl{https://arxiv.org/abs/#1}}}

\bibitem[{Allan(1966)}]{1446564}
Allan, D. 1966, Proceedings of the IEEE, 54, 221,
  \dodoi{10.1109/PROC.1966.4634}

\bibitem[{Allan(1987)}]{1539968}
---. 1987, IEEE Transactions on Ultrasonics, Ferroelectrics, and Frequency
  Control, 34, 647, \dodoi{10.1109/T-UFFC.1987.26997}

\bibitem[{{Bandura} {et~al.}(2016{\natexlab{a}}){Bandura}, {Bender}, {Cliche},
  {de Haan}, {Dobbs}, {Gilbert}, {Griffin}, \& {Hsyu}}]{2016JAI.....541005B}
{Bandura}, K., {Bender}, A.~N., {Cliche}, J.~F., {et~al.} 2016{\natexlab{a}},
  Journal of Astronomical Instrumentation, 5, 1641005,
  \dodoi{10.1142/S2251171716410051}

\bibitem[{{Bandura} {et~al.}(2016{\natexlab{b}}){Bandura}, {Cliche}, {Dobbs},
  {Gilbert}, {Ittah}, {Mena Parra}, \& {Smecher}}]{2016JAI.....541004B}
{Bandura}, K., {Cliche}, J.~F., {Dobbs}, M.~A., {et~al.} 2016{\natexlab{b}},
  Journal of Astronomical Instrumentation, 5, 1641004,
  \dodoi{10.1142/S225117171641004X}

\bibitem[{Bandura {et~al.}(2014)Bandura, Addison, Amiri, Bond, Campbell-Wilson,
  Connor, Cliche, Davis, Deng, Denman, Dobbs, Fandino, Gibbs, Gilbert, Halpern,
  Hanna, Hincks, Hinshaw, Höfer, Klages, Landecker, Masui, Parra, Newburgh,
  li~Pen, Peterson, Recnik, Shaw, Sigurdson, Sitwell, Smecher, Smegal,
  Vanderlinde, \& Wiebe}]{PathfinderOverview}
Bandura, K., Addison, G.~E., Amiri, M., {et~al.} 2014, in Ground-based and
  Airborne Telescopes V, ed. L.~M. Stepp, R.~Gilmozzi, \& H.~J. Hall, Vol.
  9145, International Society for Optics and Photonics (SPIE), 738 -- 757,
  \dodoi{10.1117/12.2054950}

\bibitem[{Barnes {et~al.}(1971)Barnes, Chi, Cutler, Healey, Leeson, McGunigal,
  Mullen, Smith, Sydnor, Vessot, \& Winkler}]{5570702}
Barnes, J.~A., Chi, A.~R., Cutler, L.~S., {et~al.} 1971, IEEE Transactions on
  Instrumentation and Measurement, IM-20, 105, \dodoi{10.1109/TIM.1971.5570702}

\bibitem[{{Bhandari} {et~al.}(2018){Bhandari}, {Keane}, {Barr}, {Jameson},
  {Petroff}, {Johnston}, {Bailes}, {Bhat}, {Burgay}, {Burke-Spolaor}, {Caleb},
  {Eatough}, {Flynn}, {Green}, {Jankowski}, {Kramer}, {Krishnan}, {Morello},
  {Possenti}, {Stappers}, {Tiburzi}, {van Straten}, {Andreoni}, {Butterley},
  {Chandra}, {Cooke}, {Corongiu}, {Coward}, {Dhillon}, {Dodson}, {Hardy},
  {Howell}, {Jaroenjittichai}, {Klotz}, {Littlefair}, {Marsh}, {Mickaliger},
  {Muxlow}, {Perrodin}, {Pritchard}, {Sawangwit}, {Terai}, {Tominaga}, {Torne},
  {Totani}, {Trois}, {Turpin}, {Niino}, {Wilson}, {Albert}, {Andr{\'e}},
  {Anghinolfi}, {Anton}, {Ardid}, {Aubert}, {Avgitas}, {Baret},
  {Barrios-Mart{\'\i}}, {Basa}, {Belhorma}, {Bertin}, {Biagi}, {Bormuth},
  {Bourret}, {Bouwhuis}, {Br{\^a}nza{\c{s}}}, {Bruijn}, {Brunner}, {Busto},
  {Capone}, {Caramete}, {Carr}, {Celli}, {Moursli}, {Chiarusi}, {Circella},
  {Coelho}, {Coleiro}, {Coniglione}, {Costantini}, {Coyle}, {Creusot},
  {D{\'\i}az}, {Deschamps}, {De Bonis}, {Distefano}, {Palma}, {Domi},
  {Donzaud}, {Dornic}, {Drouhin}, {Eberl}, {Bojaddaini}, {Khayati},
  {Els{\"a}sser}, {Enzenh{\"o}fer}, {Ettahiri}, {Fassi}, {Felis}, {Fusco},
  {Gay}, {Giordano}, {Glotin}, {Gregoire}, {Gracia-Ruiz}, {Graf}, {Hallmann},
  {van Haren}, {Heijboer}, {Hello}, {Hern{\'a}ndez-Rey}, {H{\"o}{\ss}l},
  {Hofest{\"a}dt}, {Hugon}, {Illuminati}, {James}, {de Jong}, {Jongen},
  {Kadler}, {Kalekin}, {Katz}, {Kie{\ss}ling}, {Kouchner}, {Kreter},
  {Kreykenbohm}, {Kulikovskiy}, {Lachaud}, {Lahmann}, {Lef{\`e}vre}, {Leonora},
  {Loucatos}, {Marcelin}, {Margiotta}, {Marinelli}, {Mart{\'\i}nez-Mora},
  {Mele}, {Melis}, {Michael}, {Migliozzi}, {Moussa}, {Navas}, {Nezri},
  {Organokov}, {P{\v{a}}v{\v{a}}la{\c{s}}}, {Pellegrino}, {Perrina},
  {Piattelli}, {Popa}, {Pradier}, {Quinn}, {Racca}, {Riccobene},
  {S{\'a}nchez-Losa}, {Salda{\~n}a}, {Salvadori}, {Samtleben}, {Sanguineti},
  {Sapienza}, {Sch{\"u}ssler}, {Sieger}, {Spurio}, {Stolarczyk}, {Taiuti},
  {Tayalati}, {Trovato}, {Turpin}, {T{\"o}nnis}, {Vallage}, {Van Elewyck},
  {Versari}, {Vivolo}, {Vizzocca}, {Wilms}, {Zornoza}, \&
  {Z{\'u}{\~n}iga}}]{2018MNRAS.475.1427B}
{Bhandari}, S., {Keane}, E.~F., {Barr}, E.~D., {et~al.} 2018, \mnras, 475,
  1427, \dodoi{10.1093/mnras/stx3074}

\bibitem[{{Bhardwaj} {et~al.}(2021{\natexlab{a}}){Bhardwaj}, {Gaensler},
  {Kaspi}, {Landecker}, {Mckinven}, {Michilli}, {Pleunis}, {Tendulkar},
  {Andersen}, {Boyle}, {Cassanelli}, {Chawla}, {Cook}, {Dobbs}, {Fonseca},
  {Kaczmarek}, {Leung}, {Masui}, {Mnchmeyer}, {Ng}, {Rafiei-Ravandi}, {Scholz},
  {Shin}, {Smith}, {Stairs}, \& {Zwaniga}}]{Bhardwaj_M81}
{Bhardwaj}, M., {Gaensler}, B.~M., {Kaspi}, V.~M., {et~al.} 2021{\natexlab{a}},
  \apjl, 910, L18, \dodoi{10.3847/2041-8213/abeaa6}

\bibitem[{{Bhardwaj} {et~al.}(2021{\natexlab{b}}){Bhardwaj}, {Kirichenko},
  {Michilli}, {Mayya}, {Kaspi}, {Gaensler}, {Rahman}, {Tendulkar}, {Fonseca},
  {Josephy}, {Leung}, {Merryfield}, {Petroff}, {Pleunis}, {Sanghavi}, {Scholz},
  {Shin}, {Smith}, \& {Stairs}}]{2021arXiv210812122B}
{Bhardwaj}, M., {Kirichenko}, A.~Y., {Michilli}, D., {et~al.}
  2021{\natexlab{b}}, arXiv e-prints, arXiv:2108.12122.
\newblock \doarXiv{2108.12122}

\bibitem[{Cary {et~al.}(2021)Cary, Mena-Parra, Leung, Masui, Kaczmarek, \&
  Cassanelli}]{timing_pipeline_scary_2021}
Cary, S., Mena-Parra, J., Leung, C., {et~al.} 2021, Research Notes of the
  {AAS}, 5, 216, \dodoi{10.3847/2515-5172/ac289d}

\bibitem[{{Cassanelli} {et~al.}(2021){Cassanelli}, {Leung}, {Rahman},
  {Vanderlinde}, {Mena-Parra}, {Cary}, {Masui}, {Luo}, {Lin}, {Bij}, {Gill},
  {Baker}, {Bandura}, {Berger}, {Boyle}, {Brar}, {Chatterjee}, {Cubranic},
  {Dobbs}, {Fonseca}, {Good}, {Kaczmarek}, {Kaspi}, {Landecker}, {Lanman},
  {Li}, {McKee}, {Meyers}, {Michilli}, {Naidu}, {Ng}, {Patel}, {Pearlman},
  {Pen}, {Pleunis}, {Quine}, {Renard}, {Sanghavi}, {Smith}, {Stairs}, \&
  {Tendulkar}}]{2021arXiv210705659C}
{Cassanelli}, T., {Leung}, C., {Rahman}, M., {et~al.} 2021, arXiv e-prints,
  arXiv:2107.05659.
\newblock \doarXiv{2107.05659}

\bibitem[{{Chawla} {et~al.}(2021){Chawla}, {Kaspi}, {Ransom}, {Bhardwaj},
  {Boyle}, {Breitman}, {Cassanelli}, {Cubranic}, {Dong}, {Fonseca}, {Gaensler},
  {Giri}, {Josephy}, {Kaczmarek}, {Leung}, {Masui}, {Mena-Parra}, {Merryfield},
  {Michilli}, {M{\"u}nchmeyer}, {Ng}, {Patel}, {Pearlman}, {Petroff},
  {Pleunis}, {Rahman}, {Sanghavi}, {Shin}, {Smith}, {Stairs}, \&
  {Tendulkar}}]{2021arXiv210710858C}
{Chawla}, P., {Kaspi}, V.~M., {Ransom}, S.~M., {et~al.} 2021, arXiv e-prints,
  arXiv:2107.10858.
\newblock \doarXiv{2107.10858}

\bibitem[{Chen {et~al.}(1996)Chen, Langley, \& Dragert}]{wcda1996_chen}
Chen, X., Langley, R.~B., \& Dragert, H. 1996, in GPS Trends in Precise
  Terrestrial, Airborne, and Spaceborne Applications, ed. G.~Beutler, W.~G.
  Melbourne, G.~W. Hein, \& G.~Seeber (Berlin, Heidelberg: Springer Berlin
  Heidelberg), 70--74

\bibitem[{{CHIME Scientific Collaboration} {et~al.}(\noop{3002}2021, in
  prep.)}]{chime_overview_2021}
{CHIME Scientific Collaboration}, {et~al.} \noop{3002}2021, in prep.

\bibitem[{{CHIME/FRB Collaboration} {et~al.}(2018){CHIME/FRB Collaboration},
  {Amiri}, {Bandura}, {Berger}, {Bhardwaj}, {Boyce}, {Boyle}, {Brar},
  {Burhanpurkar}, {Chawla}, {Chowdhury}, {Cliche}, {Cranmer}, {Cubranic},
  {Deng}, {Denman}, {Dobbs}, {Fandino}, {Fonseca}, {Gaensler}, {Giri},
  {Gilbert}, {Good}, {Guliani}, {Halpern}, {Hinshaw}, {H{\"o}fer}, {Josephy},
  {Kaspi}, {Landecker}, {Lang}, {Liao}, {Masui}, {Mena-Parra}, {Naidu},
  {Newburgh}, {Ng}, {Patel}, {Pen}, {Pinsonneault-Marotte}, {Pleunis}, {Rafiei
  Ravandi}, {Ransom}, {Renard}, {Scholz}, {Sigurdson}, {Siegel}, {Smith},
  {Stairs}, {Tendulkar}, {Vand erlinde}, \& {Wiebe}}]{FRBSystemOverview}
{CHIME/FRB Collaboration}, {Amiri}, M., {Bandura}, K., {et~al.} 2018, \apj,
  863, 48, \dodoi{10.3847/1538-4357/aad188}

\bibitem[{{CHIME/FRB Collaboration} {et~al.}(2019{\natexlab{a}}){CHIME/FRB
  Collaboration}, {Amiri}, {Bandura}, {Bhardwaj}, {Boubel}, {Boyce}, {Boyle},
  {. Brar}, {Burhanpurkar}, {Cassanelli}, {Chawla}, {Cliche}, {Cubranic},
  {Deng}, {Denman}, {Dobbs}, {Fandino}, {Fonseca}, {Gaensler}, {Gilbert},
  {Gill}, {Giri}, {Good}, {Halpern}, {Hanna}, {Hill}, {Hinshaw}, {H{\"o}fer},
  {Josephy}, {Kaspi}, {Landecker}, {Lang}, {Lin}, {Masui}, {Mckinven},
  {Mena-Parra}, {Merryfield}, {Michilli}, {Milutinovic}, {Moatti}, {Naidu},
  {Newburgh}, {Ng}, {Patel}, {Pen}, {Pinsonneault-Marotte}, {Pleunis},
  {Rafiei-Ravandi}, {Rahman}, {Ransom}, {Renard}, {Scholz}, {Shaw}, {Siegel},
  {Smith}, {Stairs}, {Tendulkar}, {Tretyakov}, {Vanderlinde}, \& {Yadav}}]{R2}
---. 2019{\natexlab{a}}, \nat, 566, 235, \dodoi{10.1038/s41586-018-0864-x}

\bibitem[{{CHIME/FRB Collaboration} {et~al.}(2019{\natexlab{b}}){CHIME/FRB
  Collaboration}, {Andersen}, {Bandura}, {Bhardwaj}, {Boubel}, {Boyce},
  {Boyle}, {Brar}, {Cassanelli}, {Chawla}, {Cubranic}, {Deng}, {Dobbs},
  {Fandino}, {Fonseca}, {Gaensler}, {Gilbert}, {Giri}, {Good}, {Halpern},
  {Hill}, {Hinshaw}, {H{\"o}fer}, {Josephy}, {Kaspi}, {Kothes}, {Landecker},
  {Lang}, {Li}, {Lin}, {Masui}, {Mena-Parra}, {Merryfield}, {Mckinven},
  {Michilli}, {Milutinovic}, {Naidu}, {Newburgh}, {Ng}, {Patel}, {Pen},
  {Pinsonneault-Marotte}, {Pleunis}, {Rafiei-Ravandi}, {Rahman}, {Ransom},
  {Renard}, {Scholz}, {Siegel}, {Singh}, {Smith}, {Stairs}, {Tendulkar},
  {Tretyakov}, {Vanderlinde}, {Yadav}, \& {Zwaniga}}]{RN}
{CHIME/FRB Collaboration}, {Andersen}, B.~C., {Bandura}, K., {et~al.}
  2019{\natexlab{b}}, \apjl, 885, L24, \dodoi{10.3847/2041-8213/ab4a80}

\bibitem[{{CHIME/Pulsar Collaboration} {et~al.}(2020){CHIME/Pulsar
  Collaboration}, {Amiri}, {Bandura}, {Boyle}, {Brar}, {Cliche}, {Crowter},
  {Cubranic}, {Demorest}, {Denman}, {Dobbs}, {Dong}, {Fandino}, {Fonseca},
  {Good}, {Halpern}, {H{\"o}fer}, {Kaspi}, {Landecker}, {Lin}, {Masui},
  {McKee}, {Mena-Parra}, {Meyers}, {Michilli}, {Naidu}, {Newburgh}, {Ng},
  {Patel}, {Pinsonneault-Marotte}, {Ransom}, {Renard}, {Scholz}, {Shaw},
  {Sikora}, {Stairs}, {Tan}, {Tendulkar}, {Tretyakov}, {Vanderlinde}, {Wang},
  \& {Wang}}]{2020arXiv200805681C}
{CHIME/Pulsar Collaboration}, {Amiri}, M., {Bandura}, K.~M., {et~al.} 2020,
  arXiv e-prints, arXiv:2008.05681.
\newblock \doarXiv{2008.05681}

\bibitem[{{Denman} {et~al.}(2020){Denman}, {Renard}, {Vanderlinde}, {Berger},
  {Masui}, \& {Tretyakov}}]{2020JAI.....950014D}
{Denman}, N., {Renard}, A., {Vanderlinde}, K., {et~al.} 2020, Journal of
  Astronomical Instrumentation, 9, 2050014, \dodoi{10.1142/S2251171720500142}

\bibitem[{Duval {et~al.}(1996)Duval, Héroux, \& Beck}]{cacs1996_duval}
Duval, R., Héroux, P., \& Beck, N. 1996, GIS Conference, Vancouver, Canada,
  March 1996, 53, 7

\bibitem[{{Event Horizon Telescope Collaboration} {et~al.}(2019){Event Horizon
  Telescope Collaboration}, {Akiyama}, {Alberdi}, {Alef}, {Asada}, {Azulay},
  {Baczko}, {Ball}, {Balokovi{\'c}}, {Barrett}, {Bintley}, {Blackburn},
  {Boland}, {Bouman}, {Bower}, {Bremer}, {Brinkerink}, {Brissenden}, {Britzen},
  {Broderick}, {Broguiere}, {Bronzwaer}, {Byun}, {Carlstrom}, {Chael}, {Chan},
  {Chatterjee}, {Chatterjee}, {Chen}, {Chen}, {Cho}, {Christian}, {Conway},
  {Cordes}, {Crew}, {Cui}, {Davelaar}, {De Laurentis}, {Deane}, {Dempsey},
  {Desvignes}, {Dexter}, {Doeleman}, {Eatough}, {Falcke}, {Fish}, {Fomalont},
  {Fraga-Encinas}, {Friberg}, {Fromm}, {G{\'o}mez}, {Galison}, {Gammie},
  {Garc{\'\i}a}, {Gentaz}, {Georgiev}, {Goddi}, {Gold}, {Gu}, {Gurwell},
  {Hada}, {Hecht}, {Hesper}, {Ho}, {Ho}, {Honma}, {Huang}, {Huang}, {Hughes},
  {Ikeda}, {Inoue}, {Issaoun}, {James}, {Jannuzi}, {Janssen}, {Jeter}, {Jiang},
  {Johnson}, {Jorstad}, {Jung}, {Karami}, {Karuppusamy}, {Kawashima},
  {Keating}, {Kettenis}, {Kim}, {Kim}, {Kim}, {Kino}, {Koay}, {Koch}, {Koyama},
  {Kramer}, {Kramer}, {Krichbaum}, {Kuo}, {Lauer}, {Lee}, {Li}, {Li},
  {Lindqvist}, {Liu}, {Liuzzo}, {Lo}, {Lobanov}, {Loinard}, {Lonsdale}, {Lu},
  {MacDonald}, {Mao}, {Markoff}, {Marrone}, {Marscher}, {Mart{\'\i}-Vidal},
  {Matsushita}, {Matthews}, {Medeiros}, {Menten}, {Mizuno}, {Mizuno}, {Moran},
  {Moriyama}, {Moscibrodzka}, {M{\"u}ller}, {Nagai}, {Nagar}, {Nakamura},
  {Narayan}, {Narayanan}, {Natarajan}, {Neri}, {Ni}, {Noutsos}, {Okino},
  {Olivares}, {Ortiz-Le{\'o}n}, {Oyama}, {{\"O}zel}, {Palumbo}, {Patel}, {Pen},
  {Pesce}, {Pi{\'e}tu}, {Plambeck}, {PopStefanija}, {Porth}, {Prather},
  {Preciado-L{\'o}pez}, {Psaltis}, {Pu}, {Ramakrishnan}, {Rao}, {Rawlings},
  {Raymond}, {Rezzolla}, {Ripperda}, {Roelofs}, {Rogers}, {Ros}, {Rose},
  {Roshanineshat}, {Rottmann}, {Roy}, {Ruszczyk}, {Ryan}, {Rygl},
  {S{\'a}nchez}, {S{\'a}nchez-Arguelles}, {Sasada}, {Savolainen}, {Schloerb},
  {Schuster}, {Shao}, {Shen}, {Small}, {Sohn}, {SooHoo}, {Tazaki}, {Tiede},
  {Tilanus}, {Titus}, {Toma}, {Torne}, {Trent}, {Trippe}, {Tsuda}, {van
  Bemmel}, {van Langevelde}, {van Rossum}, {Wagner}, {Wardle}, {Weintroub},
  {Wex}, {Wharton}, {Wielgus}, {Wong}, {Wu}, {Young}, {Young}, {Younsi},
  {Yuan}, {Yuan}, {Zensus}, {Zhao}, {Zhao}, {Zhu}, {Algaba}, {Allardi},
  {Amestica}, {Bach}, {Beaudoin}, {Benson}, {Berthold}, {Blanchard},
  {Blundell}, {Bustamente}, {Cappallo}, {Castillo-Dom{\'\i}nguez}, {Chang},
  {Chang}, {Chang}, {Chen}, {Chilson}, {Chuter}, {C{\'o}rdova Rosado},
  {Coulson}, {Crawford}, {Crowley}, {David}, {Derome}, {Dexter}, {Dornbusch},
  {Dudevoir}, {Dzib}, {Eckert}, {Erickson}, {Everett}, {Faber}, {Farah},
  {Fath}, {Folkers}, {Forbes}, {Freund}, {G{\'o}mez-Ruiz}, {Gale}, {Gao},
  {Geertsema}, {Graham}, {Greer}, {Grosslein}, {Gueth}, {Halverson}, {Han},
  {Han}, {Hao}, {Hasegawa}, {Henning}, {Hern{\'a}ndez-G{\'o}mez},
  {Herrero-Illana}, {Heyminck}, {Hirota}, {Hoge}, {Huang}, {Impellizzeri},
  {Jiang}, {Kamble}, {Keisler}, {Kimura}, {Kono}, {Kubo}, {Kuroda}, {Lacasse},
  {Laing}, {Leitch}, {Li}, {Lin}, {Liu}, {Liu}, {Lu}, {Marson},
  {Martin-Cocher}, {Massingill}, {Matulonis}, {McColl}, {McWhirter}, {Messias},
  {Meyer-Zhao}, {Michalik}, {Monta{\~n}a}, {Montgomerie}, {Mora-Klein},
  {Muders}, {Nadolski}, {Navarro}, {Nguyen}, {Nishioka}, {Norton}, {Nystrom},
  {Ogawa}, {Oshiro}, {Oyama}, {Padin}, {Parsons}, {Paine}, {Pe{\~n}alver},
  {Phillips}, {Poirier}, {Pradel}, {Primiani}, {Raffin}, {Rahlin}, {Reiland},
  {Risacher}, {Ruiz}, {S{\'a}ez-Mada{\'\i}n}, {Sassella}, {Schellart}, {Shaw},
  {Silva}, {Shiokawa}, {Smith}, {Snow}, {Souccar}, {Sousa}, {Sridharan},
  {Srinivasan}, {Stahm}, {Stark}, {Story}, {Timmer}, {Vertatschitsch},
  {Walther}, {Wei}, {Whitehorn}, {Whitney}, {Woody}, {Wouterloot}, {Wright},
  {Yamaguchi}, {Yu}, {Zeballos}, \& {Ziurys}}]{2019ApJ...875L...2E}
{Event Horizon Telescope Collaboration}, {Akiyama}, K., {Alberdi}, A., {et~al.}
  2019, \apjl, 875, L2, \dodoi{10.3847/2041-8213/ab0c96}

\bibitem[{{Fonseca} {et~al.}(2020){Fonseca}, {Andersen}, {Bhardwaj}, {Chawla},
  {Good}, {Josephy}, {Kaspi}, {Masui}, {Mckinven}, {Michilli}, {Pleunis},
  {Shin}, {Tendulkar}, {Bandura}, {Boyle}, {Brar}, {Cassanelli}, {Cubranic},
  {Dobbs}, {Dong}, {Gaensler}, {Hinshaw}, {Land ecker}, {Leung}, {Li}, {Lin},
  {Mena-Parra}, {Merryfield}, {Naidu}, {Ng}, {Patel}, {Pen}, {Rafiei-Ravandi},
  {Rahman}, {Ransom}, {Scholz}, {Smith}, {Stairs}, {Vanderlinde}, {Yadav}, \&
  {Zwaniga}}]{RN2}
{Fonseca}, E., {Andersen}, B.~C., {Bhardwaj}, M., {et~al.} 2020, \apjl, 891,
  L6, \dodoi{10.3847/2041-8213/ab7208}

\bibitem[{{Jackson} {et~al.}(2016){Jackson}, {Tagore}, {Deller}, {Mold{\'o}n},
  {Varenius}, {Morabito}, {Wucknitz}, {Carozzi}, {Conway}, {Drabent},
  {Kapinska}, {Orr{\`u}}, {Brentjens}, {Blaauw}, {Kuper}, {Sluman}, {Schaap},
  {Vermaas}, {Iacobelli}, {Cerrigone}, {Shulevski}, {ter Veen}, {Fallows},
  {Pizzo}, {Sipior}, {Anderson}, {Avruch}, {Bell}, {van Bemmel}, {Bentum},
  {Best}, {Bonafede}, {Breitling}, {Broderick}, {Brouw}, {Br{\"u}ggen},
  {Ciardi}, {Corstanje}, {de Gasperin}, {de Geus}, {Eisl{\"o}ffel}, {Engels},
  {Falcke}, {Garrett}, {Grie{\ss}meier}, {Gunst}, {van Haarlem}, {Heald},
  {Hoeft}, {H{\"o}randel}, {Horneffer}, {Intema}, {Juette}, {Kuniyoshi}, {van
  Leeuwen}, {Loose}, {Maat}, {McFadden}, {McKay-Bukowski}, {McKean}, {Mulcahy},
  {Munk}, {Pandey-Pommier}, {Polatidis}, {Reich}, {R{\"o}ttgering},
  {Rowlinson}, {Scaife}, {Schwarz}, {Steinmetz}, {Swinbank}, {Thoudam},
  {Toribio}, {Vermeulen}, {Vocks}, {van Weeren}, {Wise}, {Yatawatta}, \&
  {Zarka}}]{jackson2016lbcs}
{Jackson}, N., {Tagore}, A., {Deller}, A., {et~al.} 2016, \aap, 595, A86,
  \dodoi{10.1051/0004-6361/201629016}

\bibitem[{{Josephy} {et~al.}(2021){Josephy}, {Chawla}, {Curtin}, {Kaspi},
  {Bhardwaj}, {Boyle}, {Brar}, {Cassanelli}, {Fonseca}, {Gaensler}, {Leung},
  {Lin}, {Masui}, {McKinven}, {Mena-Parra}, {Michilli}, {Ng}, {Pleunis},
  {Rafiei-Ravandi}, {Rahman}, {Sanghavi}, {Scholz}, {Smith}, {Stairs},
  {Tendulkar}, \& {Zwaniga}}]{FRB_galactic_lat2021}
{Josephy}, A., {Chawla}, P., {Curtin}, A.~P., {et~al.} 2021, arXiv e-prints,
  arXiv:2106.04353.
\newblock \doarXiv{2106.04353}

\bibitem[{Kartaschoff(1979)}]{4314804}
Kartaschoff, P. 1979, IEEE Transactions on Instrumentation and Measurement, 28,
  193, \dodoi{10.1109/TIM.1979.4314804}

\bibitem[{{Leung} {et~al.}(2021){Leung}, {Mena-Parra}, {Masui}, {Bandura},
  {Bhardwaj}, {Boyle}, {Brar}, {Bruneault}, {Cassanelli}, {Cubranic},
  {Kaczmarek}, {Kaspi}, {Landecker}, {Michilli}, {Milutinovic}, {Patel},
  {Pleunis}, {Rahman}, {Renard}, {Sanghavi}, {Stairs}, {Scholz}, {Vanderlinde},
  \& {Chime/Frb Collaboration}}]{leung2020synoptic}
{Leung}, C., {Mena-Parra}, J., {Masui}, K., {et~al.} 2021, \aj, 161, 81,
  \dodoi{10.3847/1538-3881/abd174}

\bibitem[{{Lorimer} {et~al.}(2007){Lorimer}, {Bailes}, {McLaughlin},
  {Narkevic}, \& {Crawford}}]{lorimer2007bright}
{Lorimer}, D.~R., {Bailes}, M., {McLaughlin}, M.~A., {Narkevic}, D.~J., \&
  {Crawford}, F. 2007, Science, 318, 777, \dodoi{10.1126/science.1147532}

\bibitem[{{Masui} {et~al.}(2015){Masui}, {Amiri}, {Connor}, {Deng}, {Fandino},
  {H{\"o}fer}, {Halpern}, {Hanna}, {Hincks}, {Hinshaw}, {Parra}, {Newburgh},
  {Shaw}, \& {Vanderlinde}}]{2015A&C....12..181M}
{Masui}, K., {Amiri}, M., {Connor}, L., {et~al.} 2015, Astronomy and Computing,
  12, 181, \dodoi{10.1016/j.ascom.2015.07.002}

\bibitem[{{Matthews} {et~al.}(2018){Matthews}, {Crew}, {Doeleman}, {Lacasse},
  {Saez}, {Alef}, {Akiyama}, {Amestica}, {Anderson}, {Barkats}, {Baudry},
  {Brogui{\`e}re}, {Escoffier}, {Fish}, {Greenberg}, {Hecht}, {Hiriart},
  {Hirota}, {Honma}, {Ho}, {Impellizzeri}, {Inoue}, {Kohno}, {Lopez},
  {Mart{\'\i}-Vidal}, {Messias}, {Meyer-Zhao}, {Mora-Klein}, {Nagar},
  {Nishioka}, {Oyama}, {Pankratius}, {Perez}, {Phillips}, {Pradel}, {Rottmann},
  {Roy}, {Ruszczyk}, {Shillue}, {Suzuki}, \& {Treacy}}]{2018PASP..130a5002M}
{Matthews}, L.~D., {Crew}, G.~B., {Doeleman}, S.~S., {et~al.} 2018, \pasp, 130,
  015002, \dodoi{10.1088/1538-3873/aa9c3d}

\bibitem[{{Michilli} {et~al.}(2020){Michilli}, {Masui}, {Mckinven}, {Cubranic},
  {Bruneault}, {Brar}, {Patel}, {Boyle}, {Stairs}, {Renard}, {Bandura},
  {Berger}, {Breitman}, {Cassanelli}, {Dobbs}, {Kaspi}, {Leung}, {Mena-Parra},
  {Pleunis}, {Russell}, {Scholz}, {Siegel}, {Tendulkar}, \& {Vand
  erlinde}}]{michilli2020analysis}
{Michilli}, D., {Masui}, K.~W., {Mckinven}, R., {et~al.} 2020, arXiv e-prints,
  arXiv:2010.06748.
\newblock \doarXiv{2010.06748}

\bibitem[{{Mold{\'o}n} {et~al.}(2015){Mold{\'o}n}, {Deller}, {Wucknitz},
  {Jackson}, {Drabent}, {Carozzi}, {Conway}, {Kapi{\'n}ska}, {McKean},
  {Morabito}, {Varenius}, {Zarka}, {Anderson}, {Asgekar}, {Avruch}, {Bell},
  {Bentum}, {Bernardi}, {Best}, {B{\^\i}rzan}, {Bregman}, {Breitling},
  {Broderick}, {Br{\"u}ggen}, {Butcher}, {Carbone}, {Ciardi}, {de Gasperin},
  {de Geus}, {Duscha}, {Eisl{\"o}ffel}, {Engels}, {Falcke}, {Fallows},
  {Fender}, {Ferrari}, {Frieswijk}, {Garrett}, {Grie{\ss}meier}, {Gunst},
  {Hamaker}, {Hassall}, {Heald}, {Hoeft}, {Juette}, {Karastergiou},
  {Kondratiev}, {Kramer}, {Kuniyoshi}, {Kuper}, {Maat}, {Mann}, {Markoff},
  {McFadden}, {McKay-Bukowski}, {Morganti}, {Munk}, {Norden}, {Offringa},
  {Orru}, {Paas}, {Pandey-Pommier}, {Pizzo}, {Polatidis}, {Reich},
  {R{\"o}ttgering}, {Rowlinson}, {Scaife}, {Schwarz}, {Sluman}, {Smirnov},
  {Stappers}, {Steinmetz}, {Tagger}, {Tang}, {Tasse}, {Thoudam}, {Toribio},
  {Vermeulen}, {Vocks}, {van Weeren}, {White}, {Wise}, {Yatawatta}, \&
  {Zensus}}]{2015A&A...574A..73M}
{Mold{\'o}n}, J., {Deller}, A.~T., {Wucknitz}, O., {et~al.} 2015, \aap, 574,
  A73, \dodoi{10.1051/0004-6361/201425042}

\bibitem[{{Pleunis} {et~al.}(2021){Pleunis}, {Good}, {Kaspi}, {Mckinven},
  {Ransom}, {Scholz}, {Bandura}, {Bhardwaj}, {Boyle}, {Brar}, {Cassanelli},
  {Chawla}, {Fengqiu}, {Dong}, {Fonseca}, {Gaensler}, {Josephy}, {Kaczmarek},
  {Leung}, {Lin}, {Masui}, {Mena-Parra}, {Michilli}, {Ng}, {Patel},
  {Rafiei-Ravandi}, {Rahman}, {Sanghavi}, {Shin}, {Smith}, {Stairs}, \&
  {Tendulkar}}]{FRB_Morphology2021}
{Pleunis}, Z., {Good}, D.~C., {Kaspi}, V.~M., {et~al.} 2021, arXiv e-prints,
  arXiv:2106.04356.
\newblock \doarXiv{2106.04356}

\bibitem[{{Rafiei-Ravandi} {et~al.}(2021){Rafiei-Ravandi}, {Smith}, {Li},
  {Masui}, {Josephy}, {Dobbs}, {Lang}, {Bhardwaj}, {Patel}, {Bandura},
  {Berger}, {Boyle}, {Brar}, {Cassanelli}, {Chawla}, {Dong}, {Fonseca},
  {Gaensler}, {Giri}, {Good}, {Halpern}, {Kaczmarek}, {Kaspi}, {Leung}, {Lin},
  {Mena-Parra}, {Meyers}, {Michilli}, {M{\"u}nchmeyer}, {Ng}, {Petroff},
  {Pleunis}, {Rahman}, {Sanghavi}, {Scholz}, {Shin}, {Stairs}, {Tendulkar},
  {Vanderlinde}, \& {Zwaniga}}]{FRB_lss_xcor2021}
{Rafiei-Ravandi}, M., {Smith}, K.~M., {Li}, D., {et~al.} 2021, arXiv e-prints,
  arXiv:2106.04354.
\newblock \doarXiv{2106.04354}

\bibitem[{{Rogers}(1970)}]{1970RaSc....5.1239R}
{Rogers}, A. E.~E. 1970, Radio Science, 5, 1239,
  \dodoi{10.1029/RS005i010p01239}

\bibitem[{Rogers \& Moran(1981)}]{6312409}
Rogers, A. E.~E., \& Moran, J.~M. 1981, IEEE Transactions on Instrumentation
  and Measurement, IM-30, 283, \dodoi{10.1109/TIM.1981.6312409}

\bibitem[{{Schmittberger} \& {Scherer}(2020)}]{2020arXiv200409987S}
{Schmittberger}, B.~L., \& {Scherer}, D.~R. 2020, arXiv e-prints,
  arXiv:2004.09987.
\newblock \doarXiv{2004.09987}

\bibitem[{{The CHIME/FRB Collaboration} {et~al.}(2021){The CHIME/FRB
  Collaboration}, {:}, {Amiri}, {Andersen}, {Bandura}, {Berger}, {Bhardwaj},
  {Boyce}, {Boyle}, {Brar}, {Breitman}, {Cassanelli}, {Chawla}, {Chen},
  {Cliche}, {Cook}, {Cubranic}, {Curtin}, {Deng}, {Dobbs}, {Fengqiu}, {Dong},
  {Eadie}, {Fandino}, {Fonseca}, {Gaensler}, {Giri}, {Good}, {Halpern}, {Hill},
  {Hinshaw}, {Josephy}, {Kaczmarek}, {Kader}, {Kania}, {Kaspi}, {Landecker},
  {Lang}, {Leung}, {Li}, {Lin}, {Masui}, {Mckinven}, {Mena-Parra},
  {Merryfield}, {Meyers}, {Michilli}, {Milutinovic}, {Mirhosseini},
  {M{\"u}nchmeyer}, {Naidu}, {Newburgh}, {Ng}, {Patel}, {Pen}, {Petroff},
  {Pinsonneault-Marotte}, {Pleunis}, {Rafiei-Ravandi}, {Rahman}, {Ransom},
  {Renard}, {Sanghavi}, {Scholz}, {Shaw}, {Shin}, {Siegel}, {Sikora}, {Singh},
  {Smith}, {Stairs}, {Tan}, {Tendulkar}, {Vanderlinde}, {Wang}, {Wulf}, \&
  {Zwaniga}}]{CHIMEFRB_CAT1}
{The CHIME/FRB Collaboration}, {:}, {Amiri}, M., {et~al.} 2021, arXiv e-prints,
  arXiv:2106.04352.
\newblock \doarXiv{2106.04352}

\bibitem[{{Wallin} {et~al.}(2018){Wallin}, {Price}, {Carson}, \&
  {Meynadier}}]{2018ascl.soft04021W}
{Wallin}, A. E.~E., {Price}, D.~C., {Carson}, C.~G., \& {Meynadier}, F. 2018,
  {allantools: Allan deviation calculation}.
\newblock \doeprint{1804.021}

\bibitem[{{Yu} {et~al.}(2014){Yu}, {Zhang}, \& {Pen}}]{2014PhRvL.113d1303Y}
{Yu}, H.-R., {Zhang}, T.-J., \& {Pen}, U.-L. 2014, \prl, 113, 041303,
  \dodoi{10.1103/PhysRevLett.113.041303}

\end{thebibliography}

\bibliographystyle{aasjournal}



\end{document}